\documentclass{article}
\usepackage{amsfonts}

\begin{document}

\makeatletter \@addtoreset{equation}{section} \makeatother
\renewcommand{\theequation}{\thesection.\arabic{equation}}

\def\bta#1{\,\mathbb{B}_{#1}\,}
\def\btd#1{\,\overline{\mathbb{B}}_{#1}\,}

\def\etp#1{\,\mathbb{E}_{#1}\,}
\def\etn#1{\,\overline{\mathbb{E}}_{#1}\,}
\def\etpi#1{\,\mathbb{E}_{#1}^{-1}\,}
\def\etni#1{\,\overline{\mathbb{E}}_{#1}^{-1}\,}

\newcommand{\myfrac}[2]{\frac{\displaystyle #1}{\displaystyle #2}}


\title{
  Implementation of the B\"{a}cklund transformations 
  for the Ablowitz-Ladik hierarchy.
}
\author{V.E. Vekslerchik
\\[5mm]
Universidad de Castilla-La Mancha, Ciudad Real, Spain
\\
and
\\
Institute for Radiophysics and Electronics, Kharkov, Ukraine
\\[5mm]
E-mail: \texttt{vadym@ind-cr.uclm.es} }
\date{\today}
\maketitle

\begin{abstract}
The derivation of the B\"{a}cklund transformations (BTs) is a standard 
problem of the theory of the integrable systems. 
Here, I discuss the equations describing the BTs for the Ablowitz-Ladik 
hierarchy (ALH), which have been already obtained by several authors. 
The main aim of this work is to solve these equations. 
This can be done in the framework of the so-called functional 
representation of the ALH, when an infinite number of the evolutionary 
equations are replaced, using the Miwa's shifts, with a few equations 
linking tau-functions with different arguments. It is shown that starting 
from these equations it is possible to obtain explicit solutions of the BT 
equations. In other words, the main result of this work is a presentation 
of the discrete BTs as a superposition of an infinite number of evolutionary 
flows of the hierarchy. These results are used to derive the superposition 
formulae for the BTs as well as pure soliton solutions.
\end{abstract}

\section{Introduction.}

The present paper is devoted to the B\"{a}cklund transformations
(BTs) for the Ablowitz-Ladik hierarchy (ALH) \cite{AL1,AL2}. Since the
discovery of the inverse scattering transform (IST) these transformations
have become one of the most powerful tools for generating solutions for
nonlinear integrable equations (one can find the discussion of various 
aspects of this subject in the monograph \cite{RS}). 
The scheme of the solution generation can be briefly
described as follows.  One of the main points of the IST is the so-called
zero-curvature representation (ZCR), when a nonlinear equation is
presented as a compatibility condition for some overdetermined auxiliary
{\em linear} system. To this linear system one can apply the Darboux
transform.  This leads to some new solutions of the auxiliary problem
and, consequently, of the initial nonlinear equation, which are determined
(implicitly in general case) by the solutions we started with. In other
words, the Darboux transformations for the auxiliary linear equations give
us the algorithm to construct the B\"{a}cklund transformations for the
nonlinear one. This scheme was applied in 1970s to almost all known at that 
time integrable nonlinear equations (one can find the main results of these
studies in the book \cite{MS} and references therein).
This construction has been implemented by several authors for the ALH 
\cite{CM,GX,PBL,R} who obtained the systems describing the corresponding BTs.

The main aim of this work is not to derive these equations but to try to 
\textit{solve} them. To do this I will combine the standard algorithm based on 
the ZCR with another approach to the BTs, which is associated mostly with the 
KP or 2D Toda lattice equations and which is based on rewriting the 
equation in question in terms of the so-called Miwa variables 
(see below). One can find a rather detailed description of this 
technique in the paper by Adler and van Moerbeke \cite{AvM}. 
During a long time the IST and the Miwa's representation were used, in some 
sense, independently: some results have been derived using one approach, 
some using the other. 
One of the motives to write this work is 
to demonstrate that the combination of these two methods is a very 
useful strategy which gives possibility of solving some problems 
which are difficult to solve using only one of them. The BTs for 
the ALH discussed in this paper is a bright example of such 
problem.
It turns out that combining both these approaches is rather fruitful and gives 
possibility of obtaining formally \textit{explicit} expressions for the BTs 
which is the main result of this paper.

The plan of the paper is as follows.
After deriving the equations describing the BTs (section \ref{sec-der}) and 
presenting the necessary facts related to the functional representation of the 
ALH (section \ref{sec-alh}) I will obtain the formal solution of the 
B\"{a}cklund equations, i.e. present the explicit realization of the BTs for 
the ALH (section \ref{sec-rea}). These will be used to calculate the 
superposition of several BTs (section \ref{sec-sup}) and then to rederive 
the $N$-soliton solutions (section \ref{sec-sol}).

\section{Derivation of the B\"{a}cklund equations. \label{sec-der}}

All equations of the ALH, which is an infinite set of ordinal 
differential-difference equations, can be presented as the 
compatibility condition for the linear system
\begin{eqnarray}
  \Psi_{n+1} &=& U_{n} \Psi_{n}
\label{zcr-sp}\\
  \partial_{j} \Psi_{n} &=& V_{n}^{(j)} \Psi_{n} 
\label{zcr-evol}
\end{eqnarray}
where $\partial_{j}$ stands for $\partial / \partial z_{j}$, 
$z_{j}$s are an infinite set of variables (times), $\Psi_{n}$ is 
a $2$-column (or $2 \times 2$ matrix), $U_{n}$ is given by

\begin{equation}
  U_{n} = U_{n} (\lambda) = 
  \pmatrix{ \lambda & r_{n} \cr q_{n} & \lambda^{-1} } 
\label{zcr-u}
\end{equation}
and $V_{n}^{(j)}$s are some $2 \times 2$ matrices whose elements 
are some polynomials in $\lambda$, $\lambda^{-1}$ and depend on 
$q_{n}$, $r_{n}$, $q_{n \pm 1}$, $r_{n \pm 1}$, ... In what 
follows we will not use their explicit form (one can find how to 
construct these matrices in the pioneering work \cite{AL1} or, 
e.g., in the book \cite{AS}). To provide the self-consistency of 
the system (\ref{zcr-sp}), (\ref{zcr-evol}) the matrices $V_{n}$ 
have to satisfy the following equations:

\begin{equation}
  \partial_j U_{n} = V_{n+1}^{(j)} U_{n} - U_{n} V_{n}^{(j)} 
\label{ZCR}
\end{equation}
which, when rewritten in terms of $q_{n}$s and $r_{n}$s, 
constitute the ALH.

From the viewpoint of the IST, the BTs are the transformations of the form

\begin{equation}
  \widetilde\Psi_{n} = M_{n} \Psi_{n} 
\label{zcr-bd}
\end{equation}
which are compatible with the $n \to n+1$ shift. This means that 
$\widetilde\Psi_{n}$ should satisfy the equation similar to 
(\ref{zcr-sp}) with the matrix $U_{n}$ being replaced with some 
other matrix $\widetilde{U}_{n}$ of the same structure, which 
leads to the following equation for the matrix $M_{n}$:

\begin{equation}
  M_{n+1} U_{n} -  \widetilde{U}_{n} M_{n} = 0 
\label{zcr-bd-eq}
\end{equation}
where
\begin{equation}
U_{n} =
  \pmatrix{ \lambda & r_{n} \cr q_{n} & \lambda^{-1} },
\qquad
\widetilde{U}_{n} =
  \pmatrix{ \lambda & \widetilde{r}_{n} \cr 
            \widetilde{q}_{n} & \lambda^{-1} }
\end{equation}
Rewriting this matrix equation as a scalar system for the elements of $M_{n}$,

\begin{equation}
  M_{n} = \pmatrix{ a_{n} & b_{n} \cr c_{n} & d_{n} }
\end{equation}
one comes to the equations
\begin{eqnarray}
  \lambda a_{n+1} + q_{n} b_{n+1} & = & 
  \lambda a_{n} + \widetilde{r}_{n} c_{n} 
\label{bd-gen-a}
\\[2mm] 
  \lambda^{-1} d_{n+1} + r_{n} c_{n+1} & = & 
  \lambda^{-1} d_{n} + \widetilde{q}_{n} b_{n} 
\label{bd-gen-d}
\\[2mm] 
  \lambda^{-1} b_{n+1} + r_{n} a_{n+1} & = & 
  \lambda b_{n} + \widetilde{r}_{n} d_{n} 
\label{bd-gen-b}
\\[2mm] 
  \lambda c_{n+1} + q_{n} d_{n+1} & = & 
  \lambda^{-1} c_{n} + \widetilde{q}_{n} a_{n}. 
\label{bd-gen-c}
\end{eqnarray}
A simple analysis of (\ref{bd-gen-a})--(\ref{bd-gen-c}) leads to the 
conclusion that there exist two types of solutions of this system 
(compare, e.g., with the transformations of the first and second kinds 
in the paper \cite{PBL}).

\subsection{$\bta{}$-transformations.}

This type of transformations is determined by the following choice of 
the dependence of $M_{n}$ on $\lambda$:

\begin{equation}
  M_{n} = 
  \pmatrix{
    \lambda^{2} a + a'\alpha_{n}     & \lambda a r_{n} \cr
    \lambda a \widetilde{q}_{n-1} & d }. 
\end{equation}
System (\ref{bd-gen-a})--(\ref{bd-gen-c}) now becomes 
\begin{eqnarray}
&&
  a' \left( \alpha_{n+1} - \alpha_{n} \right) +
  a  \left( q_{n}r_{n+1} - \widetilde{q}_{n-1} \widetilde{r}_{n} \right)
  = 0
\\&&
  a \, r_{n+1} + a' \, r_{n}\alpha_{n+1} - d \, \widetilde{r}_{n} 
  = 0
\\&&  
  a \, \widetilde{q}_{n-1} + a' \, \widetilde{q}_{n}\alpha_{n} - d \, q_{n} 
  = 0.
\end{eqnarray}
By introducing the tau-functions of the ALH,

\begin{equation}
  q_{n} = \frac{ \sigma_{n} }{ \tau_{n} },
\qquad
  r_{n} = \frac{ \rho_{n} }{ \tau_{n} },
\qquad
  \tau_{n}^{2} = \tau_{n-1}\tau_{n+1} + \rho_{n}\sigma_{n} 
\label{tau-def}
\end{equation}
one can find $\alpha_{n}$,

\begin{equation}
  \alpha_{n} =
  \frac{ \tau_{n-1} \widetilde{\tau}_{n}   }
       { \tau_{n}   \widetilde{\tau}_{n-1} },
\end{equation}
while the remaining equations can be rewritten as 
\begin{eqnarray}
  && 
    a  \rho_{n}   \widetilde{\tau}_{n-1} 
  + a' \rho_{n-1} \widetilde{\tau}_{n} 
  - d  \tau_{n}   \widetilde{\rho}_{n-1} 
  = 0
\label{bta-eq-1}
\\
  && 
    a  \tau_{n}   \widetilde{\sigma}_{n-1} 
  + a' \tau_{n-1} \widetilde{\sigma}_{n} 
  - d  \sigma_{n} \widetilde{\tau}_{n-1} 
  = 0. 
\label{bta-eq-2}
\end{eqnarray}
These equations are enough to meet (\ref{zcr-bd-eq}). However, the condition 
  $ 
  \left(\widetilde{\tau}_{n} \right)^{2} =
  \widetilde{\tau}_{n-1}\widetilde{\tau}_{n+1} +
  \widetilde{\rho}_{n}\widetilde{\sigma}_{n}  
  $
imposes some additional restrictions on the transformed (tilded) tau-functions,
which can be presented in the form 

\begin{equation}
    a' \rho_{n}   \widetilde{\sigma}_{n} 
  - d' \tau_{n}   \widetilde{\tau}_{n} 
  + d  \tau_{n+1} \widetilde{\tau}_{n-1} 
  = 0
\label{bta-eq-3}
\end{equation}
The system (\ref{bta-eq-1})--(\ref{bta-eq-3}) completely determines this kind 
of transformations, which in what follows I will denote by the symbol 
$\bta{}$ instead of tilde, $\bta{} f = \widetilde{f}$.

\subsection{$\btd{}$-transformations.}

This type of transformations corresponds to the following choice
of the $M$-matrix:

\begin{equation}
  \overline{M}_{n}(\lambda) = 
  \pmatrix{
    \bar{a} & 
    \lambda^{-1} \bar{d} \widetilde{r}_{n-1} \cr
    \lambda^{-1} \bar{d} q_{n} & 
    \bar{d}' \bar\alpha_{n} + \lambda^{-2} \bar{d}
}
\end{equation}
In what follows all quantities related to this kind of transformations,
which I will denote by the symbol $\btd{}$, 
will be marked with overbars, which {\it does not mean complex 
conjugation}. 
After calculating $\bar\alpha_{n}$, 
  $\bar\alpha_{n} = 
   \tau_{n-1}\widetilde\tau_{n} /
   \tau_{n}  \widetilde\tau_{n-1}$
one can derive equations similar to (\ref{bta-eq-1})--(\ref{bta-eq-3}),
\begin{eqnarray}
&& 
    \bar{d}  \, \tau_{n}   \widetilde\rho_{n-1}
  + \bar{d}' \, \tau_{n-1} \widetilde\rho_{n} 
  - \bar{a}  \, \rho_{n}   \widetilde\tau_{n-1}
  = 0
\label{btd-eq-1}
\\&& 
    \bar{d}  \, \sigma_{n+1} \widetilde\tau_{n}
  + \bar{d}' \, \sigma_{n}   \widetilde\tau_{n+1} 
  - \bar{a}  \, \tau_{n+1}   \widetilde\sigma_{n}
  = 0
\label{btd-eq-2}
\\&&
    \bar{a}  \, \tau_{n+1} \widetilde\tau_{n-1}
  - \bar{a}' \, \tau_{n}   \widetilde\tau_{n} 
  + \bar{d}' \, \sigma_{n} \widetilde\rho_{n}
  = 0. 
\label{btd-eq-3}
\end{eqnarray}

Thus we have obtained equations describing two types of elementary 
B\"{a}cklund transformations:

\begin{equation}
  \bta{}: \qquad
  \left\{
  \begin{array}{l}
    a  \, \rho_{n}   \left( \bta{} \tau_{n-1} \right)
  + a' \, \rho_{n-1} \left( \bta{} \tau_{n}   \right)
  - d  \, \tau_{n}   \left( \bta{} \rho_{n-1} \right)
  = 0
  \\
    a  \, \tau_{n}   \left( \bta{} \sigma_{n-1} \right)
  + a' \, \tau_{n-1} \left( \bta{} \sigma_{n}   \right)
  - d  \, \sigma_{n} \left( \bta{} \tau_{n-1}   \right)
  = 0
  \\
    a' \, \rho_{n}   \left( \bta{} \sigma_{n} \right)
  - d' \, \tau_{n}   \left( \bta{} \tau_{n}   \right)
  + d  \, \tau_{n+1} \left( \bta{} \tau_{n-1} \right)
  = 0
  \end{array}
  \right.
\label{bta-eqs}
\end{equation}
and
\begin{equation}
  \btd{}: \qquad
  \left\{
  \begin{array}{l}
    \bar{d}  \, \tau_{n+1} \left( \btd{} \rho_{n} \right)
  + \bar{d'} \, \tau_{n}   \left( \btd{} \rho_{n+1} \right)
  - \bar{a}  \, \rho_{n+1} \left( \btd{} \tau_{n} \right)
  = 0
  \\
    \bar{d}  \, \sigma_{n+1} \left( \btd{} \tau_{n}   \right)
  + \bar{d}' \, \sigma_{n}   \left( \btd{} \tau_{n+1} \right)
  - \bar{a}  \, \tau_{n+1}   \left( \btd{} \sigma_{n} \right)
  = 0
  \\
    \bar{a}  \, \tau_{n+1} \left( \btd{} \tau_{n-1} \right)
  - \bar{a}' \, \tau_{n}   \left( \btd{} \tau_{n}   \right)
  + \bar{d}' \, \sigma_{n} \left( \btd{} \rho_{n} \right)
  = 0.
  \end{array}
  \right.
\label{btd-eqs}
\end{equation}

Writing equations (\ref{bta-eqs}) and (\ref{btd-eqs}) is a crucial 
step in constructing the BTs for the equations in question. Now one 
can, in principle, forget about the ZCR for the ALH which was our 
starting point and consider the bilinear systems (\ref{bta-eqs}), 
(\ref{btd-eqs}) as the definitions of our transformations (which could 
be derived by other methods, say, the Hirota's bilinear approach).

Usually from the practical viewpoint the BTs are used as follows. 
One takes simple solutions of the integrable system in question (the 
ALH in our case) and then tries to solve the B\"{a}cklund equations 
(equations (\ref{bta-eqs}) or (\ref{btd-eqs}) in our case) to obtain some more
complicated ones 
(probably the first who applied BTs for generation of solutions of soliton 
equations were Seeger et al \cite{SDK}). 
It is possible to solve the B\"{a}cklund equations explicitly if we start 
with some trivial solutions (e.g., constant or pure soliton), 
but to do it in general case is a task not less difficult than to 
solve the initial one. 
Nevertheless, it turns out (and it is the main point of this work) that 
it is possible to derive some (maybe formal) solutions 
for the B\"{a}cklund system (equations (\ref{bta-eqs}) or (\ref{btd-eqs})).
This is difficult to do if one restricts oneself to some particular
integrable equation (say, to the discrete nonlinear Schr\"{o}dinger 
equation or discrete modified KdV equation, both of which belong to the ALH). 
At the same time, 
information contained in \textit{all} equations of the hierarchy 
is sufficient to tackle this problem. We can construct a 
discrete (B\"{a}cklund) flow of an infinite number of continuous ones.
In other words, if we know that our tau-functions are solutions of 
the ALH (i.e. of the infinite set of compatible differential 
equations), then we can use them to derive solutions of (\ref{bta-eqs}) 
and (\ref{btd-eqs}). This will be done in section \ref{sec-rea} 
after presenting some basic facts about the ALH in the next section.

\section{Ablowitz-Ladik hierarchy. \label{sec-alh}}

Four our purposes we need the so-called functional representation 
of the ALH, which has been elaborated in \cite{V1998,V2002}, when the 
infinite set of differential-difference equations of the ALH is presented 
as a few functional equations relating tau-functions of different (shifted) 
arguments. This way of rewriting a hierarchy using the Miwa's shifts 
(in the form of the so-called Fay's identities) is very convenient 
when one studies the hierarchy 'as a whole' especially in the case 
like ours when we want to deal with \textit{all} evolutionary flows 
simultaneously. The reader can find the derivation of this form of the 
ALH together with some formulae which will be used below in \cite{V2002}. 
However, to simplify the following calculations it seems fruitful to 
rewrite the results of \cite{V2002} in a more general way. To this end 
it should be noted that some formulae of the previous section, as well 
as of \cite{V2002}, are related by the shifts 
  $\rho_{n} \to \tau_{n} \to \sigma_{n}$. 
That is why I rewrite the tau-functions as
\begin{equation}
  \rho_{n} = \tau^{-1}_{n}   
  \qquad
  \tau_{n} = \tau^{0}_{n}   
  \qquad
  \sigma_{n} = \tau^{1}_{n}
\end{equation}
and introduce a double infinite set of tau-functions
  $ \tau^{m}_{n}, \quad m=\pm 2,\pm 3,... $
using a generalization of the relation written in (\ref{tau-def}),
\begin{equation}
  \left( \tau^{m}_{n} \right)^{2} =  
  \tau^{m}_{n-1} \tau^{m}_{n+1} + \tau^{m-1}_{n} \tau^{m+1}_{n}. 
\label{alh-restr}
\end{equation}
Thus I \textit{define}
\begin{equation}
  \tau^{m}_{n} =  
  \frac{ 1 }{ \tau^{m-2}_{n} }
  \left[ 
    \left( \tau^{m-1}_{n} \right)^{2} - \tau^{m-1}_{n-1} \tau^{m-1}_{n+1}
  \right]
\qquad
  \mbox{for}
\qquad
  m = 2, 3, ...
\label{tau-pos-def}
\end{equation}
and 
\begin{equation}
  \tau^{m}_{n} =  
  \frac{ 1 }{ \tau^{m+2}_{n} }
  \left[ 
    \left( \tau^{m+1}_{n} \right)^{2} - \tau^{m+1}_{n-1} \tau^{m+1}_{n+1}
  \right]
\qquad
  \mbox{for}
\qquad
  m = -2, -3, ...
\label{tau-neg-def}
\end{equation}
It can be shown that all equations of the ALH are compatible with 
(\ref{tau-pos-def}) and (\ref{tau-neg-def}) and all equations of the extended 
version of the ALH can be obtained from the following ones:
\begin{eqnarray}
  0 & = & 
  \zeta \, \tau^{m}_{n-1} \left( \etp{\zeta}\tau^{m+1}_{n+1} \right)
  + \tau^{m+1}_{n} \left( \etp{\zeta}\tau^{m}_{n}   \right)
  - \tau^{m}_{n}   \left( \etp{\zeta}\tau^{m+1}_{n} \right)
\label{alh-pos-a}
\\
  0 & = & 
    \tau^{m}_{n-1} \left( \etp{\zeta}\tau^{m}_{n+1} \right)
  + \tau^{m-1}_{n} \left( \etp{\zeta}\tau^{m+1}_{n} \right)
  - \tau^{m}_{n}   \left( \etp{\zeta}\tau^{m}_{n}   \right)
\label{alh-pos-d}
\end{eqnarray}
(the positive subhierarchy) and
\begin{eqnarray}
  0 & = & 
  \zeta \, \tau^{m}_{n+1} \left( \etn{\zeta}\tau^{m+1}_{n-1} \right)
  + \tau^{m+1}_{n} \left( \etn{\zeta}\tau^{m}_{n}   \right)
  - \tau^{m}_{n}   \left( \etn{\zeta}\tau^{m+1}_{n} \right)
\label{alh-neg-a}
\\
  0 & = & 
    \tau^{m}_{n+1} \left( \etn{\zeta}\tau^{m}_{n-1} \right)
  + \tau^{m-1}_{n} \left( \etn{\zeta}\tau^{m+1}_{n} \right)
  - \tau^{m}_{n}   \left( \etn{\zeta}\tau^{m}_{n}   \right)
\label{alh-neg-d}
\end{eqnarray}
(the negative one).
Here the symbols $\etp{\xi}$ and $\etn{\eta}$ stand for the Miwa's shifts 
\begin{equation}
  \etp{\zeta} f( z, \bar{z} ) = f( z + i[\zeta], \bar{z} ),
\qquad
  \etn{\zeta} f( z, \bar{z} ) = f( z, \bar{z} + i[\zeta] )
\end{equation}
which are defined for functions of an infinite number of variables, 
\begin{equation}
  f(z, \bar{z}) =
  f \left(
    z_{1}, z_{2}, z_{3}, \dots
    \bar{z}_{1}, \bar{z}_{2}, \bar{z}_{3}, \dots
\right),
\end{equation}
by
\begin{eqnarray}
  f(z + \alpha [\xi], \bar{z} + \beta [\eta])
  & = &
  f(
  z_{1} + \alpha \xi,
  z_{2} + \alpha \xi^{2}/2,
  z_{3} + \alpha \xi^{3}/3,
  \dots
\cr&&
  \phantom{f(}
  \bar{z}_{1} + \beta \eta,
  \bar{z}_{2} + \beta \eta^{2}/2,
  \bar{z}_{3} + \beta \eta^{3}/3,
  \dots
  \; ). 
\end{eqnarray}
The extended version of the ALH (\ref{alh-pos-a})--(\ref{alh-neg-d})
was written in the paper \cite{S} by Sadakane who showed, using 
the free fermions approach, that the ALH arises from a reduction of the 
two-component Toda lattice hierarchy, discussed the ALH in the context 
of the affine Lie algebra that acts on the universal Grassmann 
manifold and clarified the universality of the ALH.

In what follows, we also need some formulae from \cite{V2002} 
describing superposition of Miwa's shifts. They are collected 
in their extended form (i.e. in terms of the tau-functions of 
the extended ALH hierarchy, $\tau_{n}^{m}$) in the appendix.

Now we have all necessary to 'solve' equations (\ref{bta-eqs}) 
and (\ref{btd-eqs}) which can be rewritten as 
\begin{equation}
  \bta{}: \qquad
  \left\{
  \begin{array}{l}
    a  \, \tau^{m}_{n+1}   \left( \bta{} \tau^{m+1}_{n}   \right)
  + a' \, \tau^{m}_{n}     \left( \bta{} \tau^{m+1}_{n+1} \right) 
  - d  \, \tau^{m+1}_{n+1} \left( \bta{} \tau^{m}_{n}     \right) 
  = 0 
  \\[2mm]
    a' \, \tau^{m-1}_{n} \left( \bta{} \tau^{m+1}_{n} \right) 
  - d' \, \tau^{m}_{n}   \left( \bta{} \tau^{m}_{n}   \right) 
  + d  \, \tau^{m}_{n+1} \left( \bta{} \tau^{m}_{n-1} \right) 
  = 0 
  \end{array}
  \right.
\label{bta-ext}
\end{equation}
and
\begin{equation}
  \btd{}: \qquad
  \left\{
  \begin{array}{l}
    \bar{d}  \, \tau^{m+1}_{n+1} \left( \btd{} \tau^{m}_{n}   \right)
  + \bar{d'} \, \tau^{m+1}_{n}   \left( \btd{} \tau^{m}_{n+1} \right)
  - \bar{a}  \, \tau^{m}_{n+1}   \left( \btd{} \tau^{m+1}_{n} \right)
  = 0
  \\[2mm]
    \bar{a}  \, \tau^{m}_{n+1} \left( \btd{} \tau^{m}_{n-1} \right)
  - \bar{a}' \, \tau^{m}_{n}   \left( \btd{} \tau^{m}_{n}   \right)
  + \bar{d}' \, \tau^{m+1}_{n} \left( \btd{} \tau^{m-1}_{n} \right)
  = 0. 
  \end{array}
  \right.
\label{btd-ext}
\end{equation}

\section{Solution of the B\"{a}cklund equations. \label{sec-rea}}

As it was mentioned at the end of section \ref{sec-der}, one can obtain, 
starting from the fact that $\tau^{m}_{n}$ solve the ALH equations, 
some particular solutions of the B\"{a}cklund equations (\ref{bta-ext}) and 
(\ref{btd-ext}). Indeed, by multiplying (\ref{alh-pos-a}) by 
$(d/a)^{m} (d/d')^{n}$ one can come to the conclusion that the quantity

\begin{equation}
  \mathbb{B}^{(1)} \, \tau^{m}_{n} = 
  \left( \frac{d}{a}  \right)^{m} 
  \left( \frac{d}{d'} \right)^{n}
  \etp{\zeta} \tau^{m}_{n+1}
\label{bta-sol-1a}
\end{equation}
solves the first of equations (\ref{bta-ext}) provided 
\begin{equation}
  \zeta = - \frac{ a' d }{ a d' }.
\end{equation}
In a similar way, one can check that (\ref{bta-sol-1a}) also satisfies the 
second one. Thus, we have derived some BT which can be expressed in terms 
of the Miwa's shifts (i.e. the evolutionary flows) accompanied by the shifts 
of the site index $n$. Of course, the transform $\mathbb{B}^{(1)}$ is almost 
trivial and is not interesting from the viewpoint of applications. 
However, this transformation is not the only one and it is possible to 
derive three other similar transformations, involving both positive and 
negative shifts: 
\begin{eqnarray}
  \mathbb{B}^{(1)} \, \tau^{m}_{n} & = & 
  \left( \myfrac{d}{a}  \right)^{m} 
  \left( \myfrac{d}{d'} \right)^{n}
  \etp{\zeta} \tau^{m}_{n+1}
\label{bta-sol-1}
  \\ 
  \mathbb{B}^{(2)} \, \tau^{m}_{n} & = & 
  \left( \myfrac{d'}{a'}  \right)^{m} 
  \left( -\myfrac{a}{a'} \right)^{n}
  \etpi{\zeta} \tau^{m-1}_{n}
\label{bta-sol-2}
  \\ 
  \mathbb{B}^{(3)} \, \tau^{m}_{n} & = & 
  \left( \myfrac{d'}{a'}  \right)^{m} 
  \left( \myfrac{d}{d'} \right)^{n}
  \etn{1/\zeta} \tau^{m}_{n}
\label{bta-sol-3}
  \\ 
  \mathbb{B}^{(4)} \, \tau^{m}_{n} & = & 
  \left( \myfrac{d}{a}  \right)^{m} 
  \left( -\myfrac{a}{a'} \right)^{n}
  \etni{1/\zeta} \tau^{m-1}_{n+1}
\label{bta-sol-4}
\end{eqnarray}
with
\begin{equation}
  \zeta = - \myfrac{ a' d }{ a d' }
\end{equation}
which can be used to construct more rich ones,
\begin{equation}
  \bta{} \tau^{m}_{n} = 
  \sum_{k=1}^{4} u^{(k)} \, \mathbb{B}^{(k)} \tau^{m}_{n}. 
\label{bta-sol-u}
\end{equation}
It is an important moment. Each of 
  $\mathbb{B}^{(k)} \tau^{m}_{n}$
solves the \textit{bilinear} ALH equations 
(\ref{alh-pos-a})--(\ref{alh-neg-d}). 
However their linear combination does not satisfy them automatically. 
Nevertheless, the integrable bilinear equations possess some peculiar features 
and it turns out that by imposing some restrictions on the functions $u^{(k)}$ 
in (\ref{bta-sol-u}) one can make $\bta{} \tau^{m}_{n}$ to solve the ALH 
equations. To do this, one needs to study the properties of the products of 
the Miwa's shifts and the trivial BTs $\mathbb{B}^{(k)}$.

From the superposition formulae for the Miwa's shifts presented in the appendix 
one can deduce that all $\mathbb{B}^{(k)}$ satisfy the similar identities:
\begin{equation}
  \lambda^{(k)}(\xi) 
  \tau^{m}_{n} 
  \left( \etp{\xi} \mathbb{B}^{(k)} \tau^{m}_{n} \right) 
  = 
  \left( \etp{\xi} \tau^{m}_{n} \right)
  \left( \mathbb{B}^{(k)} \tau^{m}_{n} \right)
  + 
  \xi \; \frac{ a }{ a' }
  \left( \etp{\xi} \tau^{m}_{n+1} \right)
  \left( \mathbb{B}^{(k)} \tau^{m}_{n-1} \right)
\end{equation}
where
\begin{eqnarray}
  \lambda^{(1)}(\xi) & = & a(\zeta,\xi) / \zeta 
  \\
  \lambda^{(2)}(\xi) & = & c(\xi,\zeta)
  \\
  \lambda^{(3)}(\xi) & = & e(\xi, 1/\zeta)
  \\
  \lambda^{(4)}(\xi) & = & g(\xi, 1/\zeta). 
\end{eqnarray}
Thus, if one takes the functions $u^{(k)}$ depending on the variables $z_{j}$ 
according to the rule
\begin{equation}
  \frac{\etp{\xi} u^{(k)} }{ u^{(k)}  } =
  \frac{ \lambda^{(k)}(\xi) }{ \lambda(\xi) } 
\label{bta-uk-pos}
\end{equation}
where $\lambda(\xi)$ is an arbitrary function satisfying the restriction 
$\lambda(0)=1$, then $\bta{} \tau^{m}_{n}$ satisfies
\begin{equation}
  \lambda(\xi) 
  \etp{\xi} \left( \bta{} \tau^{m}_{n} \right) 
  = 
  \frac{ 1 }{ \tau^{m}_{n} }
  \left[
    \left( \etp{\xi} \tau^{m}_{n} \right)
    \left( \bta{} \tau^{m}_{n} \right)
    + 
    \xi \; \frac{ a }{ a' }
    \left( \etp{\xi} \tau^{m}_{n+1} \right)
    \left( \bta{} \tau^{m}_{n-1} \right)
  \right].
\end{equation}
Now by straightforward algebra one can show that the transformation
\begin{equation}
  \widetilde\tau^{m}_{n} 
  \to
  \frac{ 1 }{ \lambda(\xi) \tau^{m}_{n} }
  \left[
    \left( \etp{\xi} \tau^{m}_{n} \right)
    \widetilde\tau^{m}_{n}
    + 
    \xi \; \frac{ a }{ a' }
    \left( \etp{\xi} \tau^{m}_{n+1} \right)
    \widetilde\tau^{m}_{n-1} 
  \right]
\label{etp-bta-tr}
\end{equation}
is compatible with the evolutionary equations: if one substitutes in, 
e.g., (\ref{alh-pos-a}) $\tau^{m}_{n}$ with $\widetilde\tau^{m}_{n}$ 
(which is a solution of the B\"{a}cklund equations (\ref{bta-ext})) and 
$\etp{\xi}\tau^{m}_{n}$ with the r.h.s of (\ref{etp-bta-tr}) then the 
resulting expression will be identically zero.

Analogously, $\bta{}\tau^{m}_{n}$ solve equations of the negative 
subhierarchy, if the dependence of the functions $u_{k}$ on the variables 
$\bar{z}_{j}$ is given by
\begin{equation}
  \frac{\etn{\eta} u^{(k)} }{ u^{(k)}  } =
  \frac{ \mu^{(k)}(\eta) }{ \mu(\eta) } 
\label{bta-uk-neg}
\end{equation}
where
\begin{eqnarray} 
  \mu^{(1)}(\eta) & = & e(\zeta, \eta)
  \\[2mm]
  \mu^{(2)}(\eta) & = & g(\zeta, \eta)
  \\[2mm]
  \mu^{(3)}(\eta) & = & \zeta \bar{a}(1/\zeta, \eta)
  \\[2mm]
  \mu^{(4)}(\eta) & = & \bar{c}(1/\zeta, \eta)
\end{eqnarray}
and $\mu(\eta)$ is an arbitrary function with $\mu(0)=1$.

Equations (\ref{bta-uk-pos}) and (\ref{bta-uk-neg}) can be easily solved 
by the ansatz
\begin{equation}
  u^{(k)}(z, \bar{z}) =
  \exp\left\{
    \sum_{j=1}^{\infty} \;
    \omega_{kj} z_{j} + \bar\omega_{kj} \bar{z}_{j}
  \right\}
\end{equation}
where
\begin{equation}
  \sum_{j=1}^{\infty} 
    \omega_{kj} \frac{\xi^{j}}{j} = 
  - i \ln \frac{ \lambda^{(k)}(\xi) }{ \lambda(\xi) } 
\end{equation}
and
\begin{equation}
  \sum_{j=1}^{\infty} 
    \bar\omega_{kj} \frac{\eta^{j}}{j} = 
  - i \ln \frac{ \mu^{(k)}(\eta) }{ \mu(\eta) }, 
\end{equation}
i.e., $\omega_{kj}$ and $\bar\omega_{kj}$ are the Taylor coefficients 
for the logarithmic derivatives of the functions 
$\lambda^{(k)}/\lambda$ and $\mu^{(k)}/\mu$:
\begin{eqnarray}
  \sum_{j=1}^{\infty} \omega_{kj} \xi^{j-1} & = & 
  - i \frac{d}{d\xi} 
    \ln \frac{ \lambda^{(k)}(\xi) }{ \lambda(\xi) } 
\\
  \sum_{j=1}^{\infty} \bar\omega_{kj} \eta^{j-1} & = & 
  - i \frac{d}{d\eta} 
    \ln \frac{ \mu^{(k)}(\eta) }{ \mu(\eta) } 
\end{eqnarray}
These formulae are nothing but the dispersion laws for the given boundary 
conditions (which determine the quantities $a(\xi,\eta)$, $e(\xi,\eta)$, 
etc, and hence all the functions $\lambda^{(k)}(\xi)$ and $\mu^{(k)}(\eta)$).

Similar algorithm can be developed for the $\btd{}$ transformations. 
Starting from almost trivial ones,
\begin{eqnarray}
  \overline{\mathbb{B}}^{(1)} \, \tau^{m}_{n} & = & 
  \left( \myfrac{ \bar{d} }{ \bar{a} }  \right)^{m} 
  \left( \myfrac{ \bar{a} }{ \bar{a}'}  \right)^{n}
  \etni{\bar\zeta} \tau^{m}_{n+1}
\label{btd-sol-1}
  \\
  \overline{\mathbb{B}}^{(2)} \, \tau^{m}_{n} & = & 
  \left( \myfrac{ \bar{d}' }{ \bar{a}' }  \right)^{m} 
  \left( -\myfrac{ \bar{d} }{ \bar{d}'}  \right)^{n}
  \etn{\bar\zeta} \tau^{m+1}_{n}
\label{btd-sol-2}
  \\
  \overline{\mathbb{B}}^{(3)} \, \tau^{m}_{n} & = & 
  \left( \myfrac{ \bar{d}' }{ \bar{a}' }  \right)^{m} 
  \left( \myfrac{ \bar{a} }{ \bar{a}'}  \right)^{n}
  \etpi{1/\bar\zeta} \tau^{m}_{n}
\label{btd-sol-3}
  \\
  \overline{\mathbb{B}}^{(4)} \, \tau^{m}_{n} & = & 
  \left( \myfrac{ \bar{d} }{ \bar{a} }  \right)^{m} 
  \left( -\myfrac{ \bar{d} }{ \bar{d}'}  \right)^{n}
  \etp{1/\bar\zeta} \tau^{m+1}_{n+1}
\label{btd-sol-4}
\end{eqnarray}
with
\begin{equation}
  \bar\zeta = - \myfrac{ \bar{a}\bar{d}' }{ \bar{a}'\bar{d} }
\end{equation}
one can construct the more general transformations
\begin{equation}
  \btd{} \tau^{m}_{n} = 
  \sum_{k=1}^{4} \bar{u}^{(k)} \, \overline{\mathbb{B}}^{(k)} \tau^{m}_{n} 
\label{btd-sol-u}
\end{equation}
where the functions $\bar{u}^{(k)}$ should be determined from the equations
\begin{equation}
  \frac{\etp{\xi} \bar{u}^{(k)} }{ \bar{u}^{(k)}  } =
  \frac{ \bar\lambda^{(k)}(\xi) }{ \bar\lambda(\xi) },
\qquad
  \frac{\etn{\eta} \bar{u}^{(k)} }{ \bar{u}^{(k)}  } =
  \frac{ \bar\mu^{(k)}(\eta) }{ \bar\mu^(\eta) } 
\label{btd-uk}
\end{equation}
with
\begin{equation}
  \begin{array}{lcl}
  \bar\lambda^{(1)}(\xi) & = & g(\xi, \bar\zeta) 
  \\
  \bar\lambda^{(2)}(\xi) & = & e(\xi, \bar\zeta)
  \\
  \bar\lambda^{(3)}(\xi) & = & c(\xi,1/\bar\zeta)
  \\
  \bar\lambda^{(4)}(\xi) & = &  \bar\zeta a(1/\bar\zeta,\xi) 
  \end{array}
\hspace{20mm}
  \begin{array}{lcl}
  \bar\mu^{(1)}(\eta) & = & \bar{c}(\eta, \bar\zeta)
  \\
  \bar\mu^{(2)}(\eta) & = & \bar{a}(\bar\zeta, \eta) / \bar\zeta
  \\
  \bar\mu^{(3)}(\eta) & = & g(1/\bar\zeta, \eta)
  \\
  \bar\mu^{(4)}(\eta) & = & e(1/\bar\zeta, \eta) 
  \end{array}
\end{equation}
and arbitrary $\bar\lambda(\xi)$ and $\bar\mu(\eta)$ satisfying 
$\bar\lambda(0) = \bar\mu(0) = 1$. 

So, we have derived the main result of this paper: we have \textit{explicitly} 
constructed the BTs for the ALH. This was done by combining the 
elementary transformations
$n \to n \pm 1$, $m \to m \pm 1$, the Miwa's shifts and the multiplication 
by linear in $n$ and $m$ exponents. Each of these transformations is 
trivial and does not change the structure of solutions. However their 
combination can produce more rich BTs which give us possibility of obtaining 
more complicated solutions from simple ones. I will return to this point in 
the conclusion and discuss it using as an example the soliton-adding 
transformations of section \ref{sec-sol}. An important moment is that to 
obtain physically interesting transformations one has to use superposition 
of $\mathbb{B}^{(k)}$ and $\overline{\mathbb{B}}^{(k)}$ with different 
$k$s. Moreover, one cannot restrict oneself with $k=1,2$ or $k=3,4$ only. 
Because of this it is impossible to obtain their infinitesimal versions 
by taking, say, the $\zeta \to 0$ and $\bar\zeta \to 0$ limits. That is an 
illustration of the discrete character of $\bta{}$ and $\btd{}$, 
contrary to the evolutionary operators $\etp{}$ and $\etn{}$.

\section{Superposition of BTs. \label{sec-sup}}

In the previous section we have derived the BTs, 
the transformations which commute with 
the $n \to n+1$ shifts as well as with the evolutionary flows. Now our aim is 
to ensure the commutativity of these transformations. It turns out that this 
problem is, in some sense, more difficult than the previous ones. The 
commutativity with the $n \to n+1$ shifts is due to the construction 
(\ref{zcr-bd-eq}): all transformations (\ref{bta-sol-1})--(\ref{bta-sol-4}) 
and (\ref{btd-sol-1})--(\ref{btd-sol-4}) automatically satisfy 
(\ref{bta-eqs}) and (\ref{btd-eqs}) which was derived from (\ref{zcr-bd-eq}). 
The commutativity with the evolutionary flows was achieved by proper choice of 
the dependence of the functions 
$u^{(k)}$ and $\bar{u}^{(k)}$ ($k=1,...,4$) on 
$z_{j}$ and $\bar{z}_{j}$ ($j=1,2,...$) which is described by equations 
(\ref{bta-uk-pos}), (\ref{bta-uk-neg}) and (\ref{btd-uk}). 

Before proceeding further I have to introduce the following notation.
Since now we will deal with different BTs, which will be distinguished by the 
additional subscript. The symbol $\mathbb{B}_{j}^{(k)}$ means the $k$th 
elementary transform (one of (\ref{bta-sol-1})--(\ref{bta-sol-4})) constructed 
using the set of parameters $a_{j}$, $a'_{j}$, $d_{j}$ and $d'_{j}$. 
Consequently all the quantities $\zeta$,  
the functions $\lambda^{(k)}$, $u^{(k)}$, $\lambda$, and 
others also will posses this index:
$\zeta_{j} = -  a'_{j} d_{j} / a_{j} d'_{j} $,
$\lambda^{(k)}_{j}(\xi) = \lambda(\xi; a_{j},a'_{j},d_{j},d'_{j})$ etc. 
Finally $\bta{j}$ is the BT corresponding to the superposition of 
$\mathbb{B}_{j}^{(k)}$ with coefficients $u_{j}^{(k)}$. The similar notation 
will be used for the BTs of the $\btd{}$-type.

Now we can return to the superposition of BTs and I would like to start with 
an important remark. It should be noted that the 
commutativity of evolutionary flows and BTs 
has been done for $\bta{}$ and $\btd{}$ being taken as combinations 
of \textit{four} elementary transformations, 
see (\ref{bta-sol-u}) and (\ref{btd-sol-u}). 
However, as will be shown below, if we want $\bta{}$ and 
$\btd{}$ to commute between themselves, we have to restrict ourselves to 
combinations of only \textit{two} elementary BTs. Indeed, if we apply the 
$\mathbb{B}_{1}^{(k)}$ to the $\bta{2}$-transformed 
tau-function $\bta{2} \tau^{m}_{n}$, we get 
\begin{equation}
  \beta_{12}^{(k)} \; 
  \tau^{m-1}_{n} 
  \left( \mathbb{B}_{1}^{(k)}\bta{2} \tau^{m}_{n} \right) 
  = 
  \frac{ d_{2} }{ a_{2} } \; 
  \left( \mathbb{B}_{1}^{(k)} \tau^{m}_{n} \right)
  \left( \bta{2} \tau^{m-1}_{n} \right)
  -
  \frac{ d_{1} }{ a_{1} } \; 
  \left( \mathbb{B}_{1}^{(k)} \tau^{m-1}_{n} \right)
  \left( \bta{2} \tau^{m}_{n}   \right)
\label{bk-b-expl}
\end{equation}
where
\begin{eqnarray}
  \beta_{12}^{(1)} & = & 
    \myfrac{a'_{2}}{a_{2}} \; \lambda_{2}(\zeta_{1}) 
  \\[1mm]
  \beta_{12}^{(2)} & = &  
    \frac{d_{2}}{a_{2}} \; 
    \frac{ 1 - \zeta_{1} / \zeta_{2} }{ \lambda_{2}(\zeta_{1}) }
  \\[1mm]
  \beta_{12}^{(3)} & = &  
    \zeta_{1} \; \mu_{2}(1/\zeta_{1}) 
  \\[1mm]
  \beta_{12}^{(4)} & = &  
    \frac{d'_{2}}{a_{2}} \; 
    \frac{ 1 - \zeta_{2} / \zeta_{1} }{ \mu_{2}(1/\zeta_{1}) }
\end{eqnarray}
A simple analysis leads to the conclusion that it impossible to chose the 
functions $\lambda_{2}(\xi)$ and $\mu_{2}(\eta)$ 
satisfying $\lambda_{2}(0) = \mu_{2}(0) = 1$ in a way to met the condition
\begin{equation}
  \beta_{12}^{(1)} = 
  \beta_{12}^{(2)} = 
  \beta_{12}^{(3)} = 
  \beta_{12}^{(4)}  
\end{equation}
and to ensure the skew-symmetry of the left-hand side of (\ref{bk-b-expl}) 
with respect to the interchange of the subscripts $1$ and $2$. 
However there is no problem, if one restricts oneself to only two elementary 
transformations. For example, if 
\begin{equation}
  \bta{j} = 
  u_{j}^{(2)} \, \mathbb{B}_{j}^{(2)} +
  u_{j}^{(3)} \, \mathbb{B}_{j}^{(3)} 
\end{equation}
(this is the situation which corresponds to the bright soliton case, 
see also section \ref{sec-sol}) then the choice
$\lambda_{j}(\xi) = 1$, $\mu_{j}(\eta) = 1 + \eta \; d_{j}/a_{j}$
together with the restriction $a'_{j}=d'_{j}$ leads to
\begin{equation}
  \beta_{12} \; 
  \tau^{m-1}_{n} \left( \bta{12} \tau^{m}_{n} \right)
  = 
  \frac{ d_{2} }{ a_{2} } 
  \left( \bta{1} \tau^{m}_{n}   \right)
  \left( \bta{2} \tau^{m-1}_{n} \right)
  -
  \frac{ d_{1} }{ a_{1} } 
  \left( \bta{1} \tau^{m-1}_{n} \right)
  \left( \bta{2} \tau^{m}_{n}   \right)
\label{btaa-1}
\end{equation}
with $\beta_{12} = d_{2} / a_{2} - d_{1} / a_{1}$.
Hereafter, the multi index of $\bta{}$ indicates superposition of a few BTs
\begin{equation}
  \bta{ij...} \tau^{m}_{n} 
  = 
  \bta{i} \! \bta{j} \, ... \, \tau^{m}_{n} 
\end{equation}
(after we have ensured the commutativity of different BTs, this 
notation has sense, because the result does not depend on the order 
of the application of BTs).

In a similar way, for 
\begin{equation}
  \bta{j} = 
  u_{j}^{(1)} \, \mathbb{B}_{j}^{(1)} +
  u_{j}^{(3)} \, \mathbb{B}_{j}^{(3)} 
\end{equation}
(this is what one has in the dark soliton case),  the choice
$\lambda_{j}(\xi) = 1 + \xi \, a_{j}/a'_{j}$, 
$\mu_{j}(\eta) = 1 + \eta a'_{j}/a_{j}$
together with the restriction $d_{j}=d'_{j}$ leads again to (\ref{btaa-1}), 
with $\beta_{12} = a'_{2} / a_{2} - a'_{1} / a_{1}$.

To conclude this analysis I would like to note that we are at the point where 
the boundary conditions play a crucial role. They entered earlier specifying 
the dispersion laws through the quantities $a(\xi, \eta)$, etc (see the 
previous section). But now their effect is a 'qualitative' one. They select the 
structure of the BTs and impose some restrictions on their parameters.

In the following consideration I will not write explicitly the values of the 
coefficients like $\beta_{jk}$ (though we already know that in the most 
relevant cases $\beta_{jk} = \zeta_{j}-\zeta_{k}$) because the 
'physical' quantities, such as $q_{n}$, $r_{n}$ of the ALH in the original 
setting, depend on ratios of the tau-functions. So, the constant (both with 
respect to the indices $n$, $m$ and the times) multipliers ($\beta_{jk}$ and 
similar ones which appear below) are not crucial.

By applying the B\"{a}cklund equations (\ref{bta-ext}) and (\ref{btd-ext}) 
to the superposition formula (\ref{btaa-1}) derived above, one can obtain a 
large number of the equivalent ones. Some of them are given by 
\begin{eqnarray}
  \beta_{12} \; 
  \tau^{m}_{n+1} \left( \bta{12} \tau^{m}_{n} \right)
  & = & 
  \frac{ a'_{2} }{ a_{2} } 
  \left( \bta{1} \tau^{m}_{n}   \right)
  \left( \bta{2} \tau^{m}_{n+1} \right)
  -
  \frac{ a'_{1} }{ a_{1} } 
  \left( \bta{1} \tau^{m}_{n+1} \right)
  \left( \bta{2} \tau^{m}_{n}   \right)
\label{btaa-2}
\\
  \beta_{12} \; 
  \tau^{m-1}_{n+1} \left( \bta{12} \tau^{m+1}_{n} \right)
  & = & 
  \frac{ d_{1} }{ a_{1} } 
  \frac{ d'_{2} }{ a_{2} } 
  \left( \bta{1} \tau^{m}_{n}   \right)
  \left( \bta{2} \tau^{m}_{n+1} \right)
  -
  \frac{ d'_{1} }{ a_{1} } 
  \frac{ d_{2} }{ a_{2} } 
  \left( \bta{1} \tau^{m}_{n+1} \right)
  \left( \bta{2} \tau^{m}_{n}   \right).
\label{btaa-3}
\end{eqnarray}

In a similar way, one can derive formulae describing the product of the 
$\btd{}$ transformations
\begin{eqnarray}
  \bar\beta_{12} \; 
  \tau^{m+1}_{n} \left( \btd{12} \tau^{m}_{n} \right)
  & = & 
  \frac{ \bar{a}_{2} }{ \bar{d}_{2} } 
  \left( \btd{1} \tau^{m}_{n}   \right)
  \left( \btd{2} \tau^{m+1}_{n} \right)
  -
  \frac{ \bar{a}_{1} }{ \bar{d}_{1} } 
  \left( \btd{1} \tau^{m+1}_{n} \right)
  \left( \btd{2} \tau^{m}_{n}   \right)
\label{btdd-1}
\\
  \bar\beta_{12} \; 
  \tau^{m}_{n+1} \left( \btd{12} \tau^{m}_{n} \right)
  & = & 
  \frac{ \bar{d}'_{2} }{ \bar{d}_{2} } 
  \left( \btd{1} \tau^{m}_{n}   \right)
  \left( \btd{2} \tau^{m}_{n+1} \right)
  -
  \frac{ \bar{d}'_{1} }{ \bar{d}_{1} } 
  \left( \btd{1} \tau^{m}_{n+1} \right)
  \left( \btd{2} \tau^{m}_{n}   \right)
\label{btdd-2}
\\
  \bar\beta_{12} \; 
  \tau^{m+1}_{n+1} \left( \btd{12} \tau^{m-1}_{n} \right)
  & = & 
  \frac{ \bar{a}_{1} }{ \bar{d}_{1} } 
  \frac{ \bar{a}'_{2} }{ \bar{d}_{2} } 
  \left( \btd{1} \tau^{m}_{n}   \right)
  \left( \btd{2} \tau^{m}_{n+1} \right)
  -
  \frac{ \bar{a}'_{1} }{ \bar{d}_{1} } 
  \frac{ \bar{a}_{2}  }{ \bar{d}_{2} } 
  \left( \btd{1} \tau^{m}_{n+1} \right)
  \left( \btd{2} \tau^{m}_{n}   \right)
\label{btdd-3}
\end{eqnarray}
and the $\bta{}$-$\btd{}$ products,
\begin{eqnarray}
  \gamma_{12} \; 
  \tau^{m}_{n} \left( \bta{1} \btd{2} \tau^{m}_{n} \right)
  & = & 
  \left( \bta{1} \tau^{m}_{n} \right)
  \left( \btd{2} \tau^{m}_{n} \right)
  -
  \frac{ a_{1} }{ d_{1} } 
  \frac{ \bar{d}_{2} }{ \bar{a}_{2} } 
  \left( \bta{1} \tau^{m+1}_{n} \right)
  \left( \btd{2} \tau^{m-1}_{n} \right) 
\label{btad-1}
\\
  \gamma_{12} \; 
  \tau^{m-1}_{n+1} \left( \bta{1} \btd{2} \tau^{m}_{n} \right)
  & = & 
  \frac{ d'_{1} }{ d_{1} } 
  \frac{ \bar{d}_{2} }{ \bar{a}_{2} } 
  \left( \bta{1} \tau^{m}_{n+1} \right)
  \left( \btd{2} \tau^{m-1}_{n} \right)
  +
  \frac{ \bar{d}'_{2} }{ \bar{a}_{2} } 
  \left( \bta{1} \tau^{m}_{n} \right)
  \left( \btd{2} \tau^{m-1}_{n+1}   \right)
\label{btad-2}
\\
  \gamma_{12} \; 
  \tau^{m+1}_{n+1} \left( \bta{1} \btd{2} \tau^{m}_{n} \right)
  & = & 
  \frac{ a_{1} }{ d_{1} } 
  \frac{ \bar{a}'_{2} }{ \bar{a}_{2} } 
  \left( \bta{1} \tau^{m+1}_{n} \right)
  \left( \btd{2} \tau^{m}_{n+1} \right)
  +
  \frac{ a'_{1} }{ d_{1} } 
  \left( \bta{1} \tau^{m+1}_{n+1} \right)
  \left( \btd{2} \tau^{m}_{n}   \right). 
\label{btad-3}
\end{eqnarray}
Here, again, $\bar\beta_{12}$ and $\gamma_{12}$ are some constants which 
depend on the boundary conditions and their explicit form should be 
established after specifying the class of solutions we are dealing with.

After we have obtained explicit formulae for the superposition 
of two BTs it is easy to generalize them to an arbitrary number of BTs 
involved by noting their determinant structure. For example, equation 
(\ref{btaa-2}) can be rewritten as 

\begin{equation}
  \bta{12} \tau^{m}_{n} =
  \frac{ 1 }{ \beta_{12} } \; 
  \frac{ 1 }{ \tau^{m}_{n+1} }
  \det\left|
  \matrix{
  \left( \bta{1} \tau^{m}_{n} \right) &  
  \left( \bta{2} \tau^{m}_{n} \right) \cr
  \frac{ a'_{1} }{ a_{1} } \left( \bta{1} \tau^{m}_{n+1} \right)&
  \frac{ a'_{2} }{ a_{2} } \left( \bta{2} \tau^{m}_{n+1} \right) 
  }
  \right|
\end{equation}
and generalized as 
\begin{equation}
  \tau^{m}_{n}[N] =
  \frac{ \Gamma[N] }{ \prod_{j=1}^{N-1} \tau^{m}_{n+j} } \;
  \det\left|
  \quad
    \left( \frac{ a'_{k} }{ a_{k} } \right)^{j-1}
    \left( \bta{k} \tau^{m}_{n+j-1} \right)
  \quad
  \right|_{j,k =  1, ... , N}
\label{bt-aa-x}
\end{equation}
where the traditional designation $\tau^{m}_{n}[N]$ is used for the $N$ times 
transformed tau-function,
\begin{equation}
  \tau^{m}_{n}[N] =
  \bta{1} ... \bta{N} \tau^{m}_{n}
\end{equation}
and the constant $\Gamma[N]$ is given by 
\begin{equation}
  \Gamma[N] = 
  \left[ \prod_{ 1 \le j < k \le N } \beta_{jk} \right]^{-1}.
\end{equation}
A proof of this formula can be given as follows.
Subtracting from the $j$th column, $j=1,...,N-1$ the $(j+1)$th one 
multiplied by the factor
$\left( a_{N} / a'_{N} \right)
 \left( \bta{N} \tau^{m}_{n+j-1} / \bta{N} \tau^{m}_{n+j} \right)$
one can make all elements of the $N$th row equal to zero, except the 
element at the $(N,N)$th place. 
In such a way the $N$th order determinant at the right-hand side 
of (\ref{bt-aa-x}) is reduced to the one of order $N-1$. 
The elements of the resulting determinant can be shown to be proportional 
to $\bta{Nk} \tau^{m}_{n+j-1}$. 
So, the larger determinant can be presented 
as the result of the application of $\bta{N}$ to the smaller one. 
Repeating this step one comes to (\ref{bt-aa-x}).

In a similar way, one can obtain other formulae describing the superposition 
of $\bta{}$-transfor\-mations, $\btd{}$-transformations as well as the 'mixed' 
ones for the superposition of both $\bta{}$- and $\btd{}$-transformations.

The superposition formula for the join action of both $\bta{}$- 
and $\btd{}$-transformations can be written as
\begin{equation}
  \tau^{m}_{n}[N,\overline{N}] =
  \frac{ \Gamma[N,\overline{N}] }  
       { \tau^{\mu}_{n} 
         \left(
           \prod_{\jmath=1}^{M-1} \tau^{\mu}_{n+\jmath}
         \right)
         \left(
           \prod_{\bar\jmath=1}^{\overline{M}-1} \tau^{\mu}_{n+\bar\jmath} 
         \right)
       }
  \; \det \mathcal{M}^{\mu}_{n} 
\label{bt-super}
\end{equation}
Here 
\begin{equation}
  \tau^{m}_{n}[N,\overline{N}] =
  \bta{1} ... \bta{N} 
  \btd{1} ... \btd{\overline{N}} \, \tau^{m}_{n},
\end{equation}
the numbers $M$ and $\overline{M}$, 
$M + \overline{M} = N + \overline{N}$, are given by
\begin{equation}
  M = N - m + \mu,
  \qquad
  \overline{M} = \overline{N} + m - \mu 
\end{equation}
and $\mathcal{M}^{\mu}_{n}$ is the 
$(N+\overline{N})\times(N+\overline{N})$ matrix
\begin{equation}
  \mathcal{M}^{\mu}_{n} =
  \left(
  \begin{array}{cc}
    \mathcal{A}^{\mu}_{n}[M,N] & 
    \mathcal{B}^{\mu}_{n}[M,\overline{N}] \\[3mm]
    \mathcal{C}^{\mu}_{n}[\overline{M},N] &
    \mathcal{D}^{\mu}_{n}[\overline{M},\overline{N}] 
  \end{array}
  \right)
\end{equation}
(the numbers in the square brackets determine the sizes of the matrices
that form the matrix $\mathcal{M}^{\mu}_{n}$: 
$\mathcal{A}^{\mu}_{n}[M,N]$ is a $M \times N$ matrix, etc).
The elements of the matrices 
$\mathcal{A}$, $\mathcal{B}$, $\mathcal{C}$ and $\mathcal{D}$ are given by 
\begin{eqnarray}
  \mathcal{A}^{\mu}_{n}[M,N]_{\jmath,\kappa} & = & 
    \left( \frac{ a'_{\kappa} }{ a_{\kappa} } \right)^{\jmath-1}
    \bta{\kappa} \tau^{\mu}_{n+\jmath-1} 
\\
  \mathcal{B}^{\mu}_{n}[M,\overline{N}]_{\jmath,\bar\kappa} & = & 
  - \frac{ \bar{d}_{\bar\kappa} }{ \bar{a}_{\bar\kappa} } 
    \left( 
      - \frac{ \bar{a}'_{\bar\kappa} }{ \bar{a}_{\bar\kappa} } 
    \right)^{\jmath-1}
    \btd{\bar\kappa} \tau^{\mu-1}_{n+\jmath-1} 
\\
  \mathcal{C}^{\mu}_{n}[\overline{M},N]_{\bar\jmath,\kappa} & = & 
  - \frac{ a_{\kappa} }{ d_{\kappa} }  
    \left( \myfrac{ d'_{\kappa} }{ d_{\kappa} } \right)^{\bar\jmath-1}
    \bta{\kappa} \tau^{\mu+1}_{n+\bar\jmath-1} 
\\
  \mathcal{D}^{\mu}_{n}[\overline{M},\overline{N}]_{\bar\jmath,\bar\kappa} 
  & = & 
    \left( 
      - \myfrac{ \bar{d}'_{\bar\kappa} }{ \bar{d}_{\bar\kappa} } 
    \right)^{\bar\jmath-1}
    \btd{\bar\kappa} \tau^{\mu}_{n+\bar\jmath-1} 
\end{eqnarray}
In all above formulae
\begin{equation}
  \jmath=1, ... , M
\qquad
  \bar\jmath=1, ... , \overline{M}
\qquad
  \kappa=1, ... , N 
\qquad
  \bar\kappa=1, ... , \overline{N} 
\end{equation}
and the factor $\Gamma[N,\overline{N}]$ is given by 
\begin{equation}
  \Gamma[N,\overline{N}] = 
  \frac{ (-)^{ (m-\mu)(N+\overline{N}) } } 
       { \beta[N] \; \gamma[N,\overline{N}] \; \bar\beta[\overline{N}] }
  \left[ 
      \left(
      \prod_{\kappa=1}^{N}  
      \frac{ d_{\kappa} }{ a_{\kappa} }
      \right)
      \left(
      \prod_{\bar\kappa=1}^{\overline{N}}  
      \frac{ \bar{d}_{\bar\kappa} }{ \bar{a}_{\bar\kappa} }
      \right)
  \right]^{m-\mu}.
\end{equation}
with 
\begin{equation}
  \beta[N] = 
    \prod_{ 1 \le \jmath < \kappa \le N } \beta_{\jmath\,\kappa} 
  \qquad
  \gamma[N,\overline{N}] = 
  \prod_{ \jmath = 1 }^{ N } 
  \prod_{ \bar\kappa = 1 }^{ \overline{N} } 
  \gamma_{ \jmath\,\bar\kappa} 
  \qquad
  \bar\beta[\overline{N}] =
  \prod_{ 1 \le \bar\jmath < \bar\kappa \le \overline{N} } 
      \bar\beta_{\bar\kappa\,\bar\jmath}. 
\end{equation}

To illustrate this general result let us consider the simplest case when 
$\mathcal{M}^{\mu}_{n}$ is a $2 \times 2$ matrix. All binary superposition 
formulae presented in this section can be obtained from (\ref{bt-super}). 
The following table contains the values of the parameters 
$N$, $\overline{N}$ and $\mu$ which lead to 
(\ref{btaa-1}), (\ref{btaa-2})--(\ref{btad-3}).

\begin{center}
\begin{tabular}{|cc|c|c|}
\hline
&&&
\\[-4mm]
  $N$ & $\overline{N}$ & \qquad $\mu$ \qquad\phantom{.} & 
  \qquad equation \qquad\phantom{.} 
\\
\hline
   &  & $m-1$ & (\ref{btaa-1})
\\
  2 & 0 & $m$ & (\ref{btaa-2})
\\
   &  & $m-2$ & (\ref{btaa-3})
\\[2mm]
\hline
   &  & $m+1$ & (\ref{btdd-1})
\\
  0 & 2 & $m$ & (\ref{btdd-2})
\\
   &  & $m+2$ & (\ref{btdd-3})
\\[2mm]
\hline
   &  & $m$ & (\ref{btad-1})
\\
  1 & 1 & $m-1$ & (\ref{btad-2})
\\
   &  & $m+1$ & (\ref{btad-3})
\\
\hline
\end{tabular}
\end{center}

\section{Solitons of the ALH. \label{sec-sol}}

In this section I will discuss two types of soliton solutions: the bright 
solitons (which correspond to the case of the zero boundary conditions) and 
the dark ones (the case of the so-called finite-density boundary conditions). 
They will be obtained by applying the $N$-fold ($2N$-fold) BTs to the vacuum 
solutions. It should be noted that the formulae of the previous section 
immediately give us the structure of $N$-soliton solutions. So, we have only 
to write the dispersion laws and to solve the question of commutativity 
of different BTs.

Contrary to traditional IST-based approach I will distinguish different types 
of solitons not by writing the boundary conditions for the functions $q_{n}$ 
and $r_{n}$, but starting from the corresponding vacuum solutions ($0$-soliton 
ones) directly in terms of the tau-functions $\tau^{m}_{n}$.

\subsection{Bright solitons of the ALH.}

The vacuum tau-function corresponding to the bright soliton case is given by
\begin{equation}
  \tau^{m}_{n} = \delta_{m,0}
\label{tau-vac-bright}
\end{equation}
or, if to return to the original tau-functions, 
$\tau_{n}$, $\sigma_{n}$ and $\rho_{n}$, by

\begin{equation}
  \tau_{n} = 1,
  \qquad
  \sigma_{n} = \rho_{n} = 0. 
\end{equation}
Applying the elementary BTs (\ref{bta-sol-1})--(\ref{bta-sol-4}), one gets 

\begin{equation}
  \begin{array}{lcl}
  \mathbb{B}^{(1)} \, \tau_{n} & = & ( d / d' )^{n} \\
  \mathbb{B}^{(2)} \, \tau_{n} & = & 0 \\
  \mathbb{B}^{(3)} \, \tau_{n} & = & ( d / d' )^{n} \\
  \mathbb{B}^{(4)} \, \tau_{n} & = & 0 
  \end{array}
\qquad
  \begin{array}{lcl}
  \mathbb{B}^{(1)} \, \sigma_{n} & = & 0 \\
  \mathbb{B}^{(2)} \, \sigma_{n} & = & ( - a / a' )^{n} \\
  \mathbb{B}^{(3)} \, \sigma_{n} & = & 0 \\
  \mathbb{B}^{(4)} \, \sigma_{n} & = & ( - a / a' )^{n}  
  \end{array}
\qquad
  \begin{array}{lcl}
  \mathbb{B}^{(1)} \, \rho_{n} & = & 0 \\
  \mathbb{B}^{(2)} \, \rho_{n} & = & 0 \\
  \mathbb{B}^{(3)} \, \rho_{n} & = & 0 \\
  \mathbb{B}^{(4)} \, \rho_{n} & = & 0 
  \end{array}
\end{equation}
It is easy to note that 
\begin{equation}
  \mathbb{B}^{(1)} =  \mathbb{B}^{(3)},
\qquad
  \mathbb{B}^{(2)} =  \mathbb{B}^{(4)}
\end{equation}
which means that general BT, $\bta{}$, can be taken as a combination of two 
elementary one, say, as
\begin{equation}
  \bta{} = 
  u^{(2)} \, \mathbb{B}^{(2)} +
  u^{(3)} \, \mathbb{B}^{(3)} 
\label{bright-bta}
\end{equation}
and to ensure the commutativity it is sufficient to impose the restriction 
\begin{equation}
  a' = d' 
\end{equation}
and to choose the functions $\lambda(\xi)$ and $\mu(\eta)$ as follows:
\begin{equation}
  \lambda(\xi) = 1,
  \qquad
  \mu(\eta) = 1 - \eta\zeta,
  \qquad
  \zeta = - \frac{ d }{ a }
\end{equation}
which leads to equations (\ref{btaa-1}), (\ref{btaa-2}) and (\ref{btaa-3}) 
with
\begin{equation}
  \beta_{12} = \zeta_{1} - \zeta_{2}.
\end{equation}
Calculating the coefficients that appear in the superposition of the 
Miwa's shifts
\begin{equation}
  a(\xi,\eta) = \bar{a}(\xi,\eta) = \xi - \eta,
  \qquad
  e(\xi,\eta) = 1 - \xi\eta 
\end{equation}
and
\begin{equation}
  c(\xi,\eta) = \bar{c}(\xi,\eta) = g(\xi,\eta) = 1 
\end{equation}
which is easy to do by the substitution of (\ref{tau-vac-bright}) in the 
formulae presented in the appendix, one comes to the following 
equations describing the dependence of the functions $u^{(k)}$ on the 
variables of the ALH:
\begin{equation}
  \begin{array}{lcl}
  \myfrac{ \etp{\xi} u^{(2)} }{ u^{(2)} } & = & 1 
  \\[2mm]
  \myfrac{ \etp{\xi} u^{(3)} }{ u^{(3)} } & = & 1 - \xi / \zeta 
  \end{array}
  \qquad
  \begin{array}{lcl}
  \myfrac{ \etn{\eta} u^{(2)} }{ u^{(2)} } & = & \myfrac{ 1}{ 1 - \eta\zeta } 
  \\[2mm]
  \myfrac{ \etn{\eta} u^{(3)} }{ u^{(3)} } & = & 1. 
  \end{array}
\label{dl-a}
\end{equation}

Similar analysis can be applied to the $\btd{}$-transformations. After 
calculating the elementary transformations 
\begin{equation}
  \begin{array}{lcl}
  \overline{\mathbb{B}}^{(1)} \tau_{n} & = & ( \bar{a} / \bar{a}' )^{n} \\
  \overline{\mathbb{B}}^{(2)} \tau_{n} & = & 0 \\
  \overline{\mathbb{B}}^{(3)} \tau_{n} & = & ( \bar{a} / \bar{a}' )^{n} \\
  \overline{\mathbb{B}}^{(4)} \tau_{n} & = & 0 
  \end{array}
\qquad
  \begin{array}{lcl}
  \overline{\mathbb{B}}^{(1)} \sigma_{n} & = & 0 \\
  \overline{\mathbb{B}}^{(2)} \sigma_{n} & = & 0 \\
  \overline{\mathbb{B}}^{(3)} \sigma_{n} & = & 0 \\
  \overline{\mathbb{B}}^{(4)} \sigma_{n} & = & 0 
  \end{array}
\qquad
  \begin{array}{lcl}
  \overline{\mathbb{B}}^{(1)} \rho_{n} & = & 0 \\
  \overline{\mathbb{B}}^{(2)} \rho_{n} & = & ( - \bar{d} / \bar{d}' )^{n} \\
  \overline{\mathbb{B}}^{(3)} \rho_{n} & = & 0 \\
  \overline{\mathbb{B}}^{(4)} \rho_{n} & = & ( - \bar{d} / \bar{d}' )^{n}  
  \end{array}
\end{equation}
and noting that 
\begin{equation}
  \overline{\mathbb{B}}^{(1)} = \overline{\mathbb{B}}^{(3)}, 
  \qquad
  \overline{\mathbb{B}}^{(2)} = \overline{\mathbb{B}}^{(4)} 
\end{equation}
one can choose
\begin{equation}
  \btd{} = 
  \bar{u}_{2} \overline{\mathbb{B}}^{(2)} + 
  \bar{u}_{3} \overline{\mathbb{B}}^{(3)}  
\label{bright-btd}
\end{equation}
which leads to the restriction
\begin{equation}
  \bar{a}' = \bar{d}'
\end{equation}
and 
\begin{equation}
  \bar\lambda(\xi) = 1, 
  \qquad
  \bar\mu(\eta) = 1 - \eta/\bar\zeta, 
  \qquad
  \bar\zeta = - \frac{ \bar{a} }{ \bar{d} }.
\end{equation}
Further, one can get $\bar\beta_{12} = \bar\zeta_{1} - \bar\zeta_{2}$ 
and the following equations:
\begin{equation}
  \begin{array}{lcl}
  \myfrac{ \etp{\xi} \bar{u}^{(2)} }{ \bar{u}^{(2)} } & = & 1 - \xi\bar\zeta
  \\[2mm]
  \myfrac{ \etp{\xi} \bar{u}^{(3)} }{ \bar{u}^{(3)} } & = & 1 
  \end{array}
\qquad
  \begin{array}{lcl}
  \myfrac{ \etn{\eta} \bar{u}^{(2)} }{ \bar{u}^{(2)} } & = & 1 
  \\[2mm]
  \myfrac{ \etn{\eta} \bar{u}^{(3)} }{ \bar{u}^{(3)} } & = &  
    \myfrac{ 1 }{ 1 - \eta/\bar\zeta }.  
  \end{array}
\label{dl-d}
\end{equation}

An important note: both $\bta{}$- and $\btd{}$-transformations, taken 
alone, destroy the symmetry between $q_{n}$ and $r_{n}$. Say,
$\bta{} q_{n} = \mbox{const}\!\cdot \zeta^{-n}$, while
$\bta{} r_{n} = 0$, which means that it is impossible to make the 
transformed functions to satisfy the 'physical' involution: 
$q_{n} = - r_{n}^{*}$ (the star stands for the complex conjugation). 
The same is valid for the $\btd{}$-transformations. Hence, to obtain 
'physical' solutions in the bright soliton case we have to use the 
$\bta{}$- and $\btd{}$-transformations only in combination, which gives 
possibility of preserving the involution. Say, to obtain the $N$-soliton 
solution we have to use the superposition of $N$ $\bta{}$- and $N$ 
$\btd{}$-transformations (with imposing some relations on parameters 
of the former and the later). Setting 
\begin{equation}
  \overline{N} = N
\end{equation}
and noting that
\begin{equation}
  \bta{\kappa} \tau_{n} = 
  \left( \frac{ d_{\kappa} }{ d'_{\kappa} } \right)^{n} 
  \bta{\kappa} \tau_{0}
\qquad
  \bta{\kappa} \sigma_{n} = 
  \left( - \frac{ a_{\kappa} }{ a'_{\kappa} } \right)^{n} 
  \bta{\kappa} \sigma_{0}
\end{equation}
and 
\begin{equation}
  \btd{\bar\kappa} \tau_{n} = 
  \left( \frac{ \bar{a}_{\bar\kappa} }{ \bar{a}'_{\bar\kappa} } \right)^{n} 
  \btd{\bar\kappa} \tau_{0}
\qquad
  \btd{\bar\kappa} \rho_{n} = 
  \left( - \frac{ \bar{d}_{\bar\kappa} }{ \bar{d}'_{\bar\kappa} } \right)^{n} 
  \btd{\bar\kappa} \rho_{0}, 
\end{equation}
one can rewrite expression (\ref{bt-super}) for the 
superposition of BTs, in the $\mu=0$ case, as
\begin{eqnarray}
  \tau^{m}_{n}[N,N] & = & 
  (-)^{\frac{1}{2} m (m+1) } \, 
  C_{*} \, 
  \zeta_{*}^{m} \, 
  \xi_{*}^{n}
\\ &&
  \times
  \det\left|
  \begin{array}{rcr} 
    \zeta_{\kappa}^{\jmath-1} 
    \left( \bta{\kappa} \tau_{0} \right)
  & \; &
    \bar\zeta_{\bar\kappa}^{-n-\jmath}
    \left( \btd{\bar\kappa} \rho_{0} \right)
  \\[5mm]
    \zeta_{\kappa}^{-n-\bar\jmath}
    \left( \bta{\kappa} \sigma_{0} \right)
  &  & 
    \bar\zeta_{\bar\kappa}^{\bar\jmath - 1}
    \left( \btd{\bar\kappa} \tau_{0} \right)
  \end{array}
  \right|_{ 
    \begin{array}{l} 
      \scriptstyle \jmath=1, ... , N-m \\[-2mm]
      \scriptstyle \bar\jmath=1, ... , N+m \\[-2mm]
      \scriptstyle \kappa,\bar\kappa=1, ... , N 
    \end{array} }
\nonumber
\end{eqnarray}
where the quantities $\zeta_{*}$, $\xi_{*}$ and $C_{*}$ are given by 

\begin{equation}
  \zeta_{*} = 
  (-)^{N} 
  \frac{ \prod_{\kappa=1}^{N} \zeta_{\kappa} }
       { \prod_{\bar\kappa=1}^{N} \bar\zeta_{\bar\kappa} }, 
  \qquad
  \xi_{*} = 
  \left(\prod_{\kappa=1}^{N} 
      \frac{ d_{\kappa} }{ d'_{\kappa} } \right)
  \left(\prod_{\bar\kappa=1}^{N}
      \frac{ \bar{a}_{\bar\kappa} }{ \bar{a}'_{\bar\kappa} } \right)
\end{equation}
and 
\begin{equation}
  C_{*} = 
  \frac{ (-)^{ \frac{1}{2}N(N-1) } } 
       { \beta[N] \gamma[N,N] \bar\beta[N] }.
\end{equation}
That leads to the following formula for the $N$-soliton solution:
\begin{equation}
  q_{n}[N,N] = 
  - \zeta_{*} 
  \frac{ \Delta^{(1)}_{n} }{ \Delta^{(0)}_{n} },
\qquad
  r_{n}[N,N] = \zeta_{*}^{-1} 
  \frac{ \Delta^{(-1)}_{n} }{ \Delta^{(0)}_{n} }
\end{equation}
where
\begin{equation}
  \Delta^{(m)}_{n} = 
  \Delta^{(m)}_{n}\left( z, \bar{z} \right) =
  \det\left|
  \begin{array}{ccc} 
    \zeta_{\kappa}^{\jmath-1} 
  & \; &
    R_{\bar\kappa}\left( z, \bar{z} \right)
    \bar\zeta_{\bar\kappa}^{-n-\jmath}
  \\[5mm]
    Q_{\kappa}\left( z, \bar{z} \right)
    \zeta_{\kappa}^{-n-\bar\jmath}
  &  & 
    \bar\zeta_{\bar\kappa}^{\bar\jmath - 1}
  \end{array}
  \right|_{ 
    \begin{array}{l} 
      \scriptstyle \jmath=1, ... , N-m \\[-2mm]
      \scriptstyle \bar\jmath=1, ... , N+m \\[-2mm]
      \scriptstyle \kappa,\bar\kappa=1, ... , N 
    \end{array} }
\end{equation}
The quantities 
  $Q_{\kappa}\left( z, \bar{z} \right)$ and 
  $R_{\bar\kappa}\left( z, \bar{z} \right)$, 
which describe the dependence on the ALH-coordi\-nates
$z_{j}$ and $\bar{z}_{j}$, 
\begin{equation}
  Q_{\kappa}\left( z, \bar{z} \right) = 
  \frac{ u^{(2)}_{\kappa} }{ u^{(3)}_{\kappa} }, 
\qquad
  R_{\bar\kappa}\left( z, \bar{z} \right) =
  \frac{ \bar{u}^{(2)}_{\bar\kappa} }{ \bar{u}^{(3)}_{\bar\kappa} }, 
\end{equation}
can be easily calculated from (\ref{dl-a}) and (\ref{dl-d}) using the 
identity
\begin{equation}
  \ln( 1 - x ) =
  - \sum_{j=1}^{\infty} 
    \frac{ x^{j} }{ j }
\end{equation}
and are given by 
\begin{eqnarray}
  Q_{\kappa}\left( z, \bar{z} \right) & = & 
  Q_{\kappa}^{(0)}
  \exp\left\{  
  - i \sum_{j=1}^{\infty} \left( 
      z_{j} \zeta_{\kappa}^{-j} + 
      \bar{z}_{j} \zeta_{\kappa}^{j} 
    \right) 
  \right\} 
\label{bright-Q}
\\
  R_{\bar\kappa}\left( z, \bar{z} \right) & = & 
  R_{\bar\kappa}^{(0)}
  \exp\left\{  
  \phantom{-} i \sum_{j=1}^{\infty} \left( 
      z_{j} \bar\zeta_{\bar\kappa}^{j} + 
      \bar{z}_{j} \bar\zeta_{\bar\kappa}^{-j} 
    \right) 
  \right\} 
\label{bright-R}
\end{eqnarray}
where $Q_{\kappa}^{(0)}$ and $R_{\bar\kappa}^{(0)}$ are some arbitrary 
constants (these expressions are nothing but the well-known 
dispersion laws for the bright solitons).

\subsection{Dark solitons of the ALH.}

The vacuum tau-function corresponding to the so-called finite-density 
boundary conditions (the dark soliton case) is given by
\begin{equation}
  \tau^{m}_{n} = 
  \alpha^{ \frac{m^{2}}{2} } \beta^{ \frac{n^{2}}{2} }
  u^{m} v^{n}
\label{dark-vacuum}
\end{equation}
where the constants $\alpha$ and $\beta$ are related by 
$\alpha + \beta = 1$, while $u$ and $v$ are some 
functions of the ALH-coordinates. Contrary to the bright solitons which are 
two parametrical, a dark soliton depends only on one parameter. 
From the viewpoint of the BTs this difference manifests itself in the 
following facts: 
(1) to construct $N$-soliton solutions we need to apply $N$ 
BTs to the vacuum solution (and not $2N$, as was in the case bright solitons) 
and 
(2) the $\bta{}$ and $\btd{}$ transformations lead to similar results. 
So, in what follows the $N$-dark soliton solutions will be obtained as a 
superposition of the $\bta{}$ transformations using (\ref{bt-aa-x}).
Thus, we have to solve now the following three problems: 
we have to ensure the commutativity of our BTs, 
to calculate the dispersion laws and to satisfy the 'physical' involution 
condition $r_{n} = q^{*}_{n}$.
As in the previous section, the first problem can be solved rather 
quickly. The case is that of four elementary BTs
(\ref{bta-sol-1})--(\ref{bta-sol-4}) only two, say 
$\mathbb{B}^{(1)}$ and $\mathbb{B}^{(3)}$, are independent while
$\mathbb{B}^{(2)} = \mathbb{B}^{(3)}$ and
$\mathbb{B}^{(4)} = \mathbb{B}^{(1)}$. 
So, one can define the $\bta{}$ transformation as 
\begin{equation}
  \bta{} = u^{(1)}\mathbb{B}^{(1)} + u^{(3)}\mathbb{B}^{(3)}. 
\end{equation}
Then it is easy to show that they will commute, if we set
\begin{equation}
  d = d'
\end{equation}
and choose 
\begin{equation}
  \lambda(\xi) = 1 - \xi / \zeta,
  \qquad
  \mu(\eta) = 1 - \zeta\eta.
\end{equation}
This leads to the superposition formula (\ref{btaa-1}) with 
$\beta_{12} = \zeta_{1} - \zeta_{2}$.

The dependence of $u^{(1,3)}$ on the ALH times is given by 
\begin{equation}
  \begin{array}{lcl}
  \myfrac{ \etp{\xi} u^{(1)} }{ u^{(1)} } & = & 
  \myfrac{ 1 }{ c(\zeta, \xi) }
  \\[3mm]
  \myfrac{ \etp{\xi} u^{(3)} }{ u^{(3)} } & = & 
  \myfrac{ 1 }{ g(\xi, 1/\zeta) }
  \end{array}
\hspace{20mm}
  \begin{array}{lcl}
  \myfrac{ \etn{\eta} u^{(1)} }{ u^{(1)} } & = & 
  \myfrac{ 1 }{ g(\zeta, \eta) }
  \\[3mm]
  \myfrac{ \etn{\eta} u^{(3)} }{ u^{(3)} } & = & 
  \myfrac{ 1 }{ \bar{c}(1/\zeta, \eta) }
  \end{array}
\label{dark-disperion}
\end{equation}
where $c$, $\bar{c}$ and $g$ are the coefficients which appear in formulae 
describing the superposition of the Miwa's shifts $\etp{}$ and $\etn{}$ 
(see appendix). 
Their explicit form, as well as the explicit form of 
the functions $u$ and $v$ from (\ref{dark-vacuum}) can obtained by 
substitution of the vacuum tau-function (\ref{dark-vacuum}) into the 
corresponding equation. This gives
\begin{eqnarray}
  c(\xi,\eta) = \bar{c}(\xi,\eta) & = &  
    \alpha 
      \left( \myfrac{\etp{\xi}u}{u}  \right)
      \left( \myfrac{\etp{\eta}u}{u} \right) +
    \beta  
      \left( \myfrac{\etp{\xi}v}{v}  \right)
      \left( \myfrac{\etp{\eta}v}{v} \right)
\\
  g(\xi,\eta) & = & 
    1 -
    \alpha\xi\eta
    \left( \myfrac{\etp{\xi}u}{u}  \right)
    \left( \myfrac{\etn{\eta}u}{u} \right)
\end{eqnarray}
and
\begin{equation}
  \begin{array}{lclcl}
  \myfrac{ \etp{\zeta} u }{ u } & = & 
  \myfrac{ \etn{\zeta} u }{ u } & = & 
  \myfrac{ 2 }{ R(\zeta) - \zeta + 1 }
\\
  \myfrac{ \etp{\zeta} v }{ v } & = & 
  \left( \myfrac{ \etn{\zeta} v }{ v } \right)^{-1} & = & 
  \myfrac{ 2 }{ R(\zeta) + \zeta + 1 }
  \end{array}
\label{vac-evol}
\end{equation}
where
\begin{equation}
  R^{2}(\zeta) = 1 + 2 (\alpha-\beta)\zeta + \zeta^{2}. 
\end{equation}
The last equations contain all necessary to obtain the dispersion laws that we 
need. However the presence of radicals $R(\zeta)$ complicates the following 
analysis, so, I would like to rewrite them using the parameterization 
\begin{equation}
  \zeta = 
  \frac{ \sin 2\theta(\zeta) }
       { \sin\left[ 2\omega - 2\theta(\zeta) \right] }
\end{equation}
where the constant parameter $\omega$ is given by
\begin{equation}
  \alpha = \cos^{2}\omega,
  \qquad
  \beta = \sin^{2}\omega.
\end{equation}
In terms of $\theta(\zeta)$,
\begin{equation}
  R(\zeta) = 
  \frac{ \sin 2\omega }
       { \sin\left[ 2\omega - 2\theta(\zeta) \right] }
\end{equation}
and
\begin{eqnarray} 
  c(\xi, \eta) & = & 
  \frac{ \cos\left[ \theta(\xi) - \theta(\eta) \right] }
       { \cos\theta(\xi) \cos\theta(\eta) }
\\
  g(\xi, \eta) & = & 
  \frac{ \sin\omega \sin\left[ \omega - \theta(\xi) - \theta(\eta) \right] }
       { \sin\left[ \omega - \theta(\xi)  \right] 
         \sin\left[ \omega - \theta(\eta) \right] }
\end{eqnarray} 
and we can present $\bta{}$ as
\begin{equation}
  \bta{}\tau^{m}_{n} = 
  \tau^{m}_{n} 
  \left( \frac{d}{a} \right)^{m}
  \left[ 
    w^{(1)} B^{(1,m)}_{n} + 
    w^{(3)} B^{(3,m)}_{n} 
  \right]
\end{equation}
with
\begin{eqnarray}
  B^{(1,m)}_{n} & = & 
    \left[ \frac{ \cos(\omega-\theta) }{ \cos\omega \cos\theta } \right]^{m}
    \left[ \frac{ \sin\omega \sin(\omega-\theta) }{ \cos\theta } \right]^{n}
\\[3mm]
  B^{(3,m)}_{n} & = & 
    \left[ \frac{ \sin(\omega-\theta) }{ \cos\omega \sin\theta } \right]^{m}
    \left[ \frac{ \sin\omega \cos(\omega-\theta) }{ \sin\theta } \right]^{n}
\end{eqnarray}
(the functions $w^{(1)}$ and $w^{(3)}$ are, up to some unimportant constants, 
$v u^{(1)}$ and $u^{(3)}$).

From these formulae, one can deduce that the condition $q_{n}=r_{n}^{*}$
can be satisfied by choosing $d/a = \cos\omega$ and introducing the 
\textit{real} parameters $\gamma_{k}$, describing the $k$th BT, by
$\theta_{k} = \omega/2 + i \gamma_{k}$, or
\begin{equation}
  \zeta_{k} =
  \frac{ \sin\left( \omega + 2i\gamma_{k} \right) }
       { \sin\left( \omega - 2i\gamma_{k} \right) }
\end{equation}
which reproduces the already known fact that the eigenvalues of the 
scattering problem, corresponding to the dark solitons are located on the 
ark of the unit circle
$ |\zeta_{k}| = 1 $, 
$ |\arg\zeta_{k}| < \pi - 2\omega $ (in our designations).
Now the BTs $\bta{k}$ can be written as
\begin{equation}
  \frac{ \bta{k}\tau^{m}_{n} }{ \tau^{m}_{n} } = 
  b_{k}
  \exp\left[ i\vartheta_{k} m + \chi_{k} n \right]
  \left\{
    w_{k} e^{ i \phi_{k} m } h_{k}^{n} +  
    w_{k}^{-1} e^{ - i \phi_{k} m } h_{k}^{-n}
  \right\}.
\end{equation}
Here, the constant $b_{k}$ is given by
$  b_{k} = 
  \sqrt{ 
    v u^{(1)}_{k} u^{(3)}_{k} \; 
    \sin\left( \frac{\omega}{2} - i\gamma_{k} \right) /
    \cos\left( \frac{\omega}{2} + i\gamma_{k} \right) 
  }
$
(I write this expression only for the sake of completeness, because $b_{k}$ 
does not appear in the final formulae) while other parameters 
(more essential) 
are given by
\begin{equation}
  \vartheta_{k} = 
  \arg \sin\left( \omega - 2i\gamma_{k} \right), 
\qquad
  \chi_{k} = 
  \ln \sin\omega + i\vartheta_{k} 
\end{equation}
and
\begin{equation}
  h_{k} = 
  \left| \tan\left( \frac{\omega}{2} + i\gamma_{k} \right) \right|, 
\qquad
  \phi_{k} = 
  \arg \tan\left( \frac{\omega}{2} + i\gamma_{k} \right). 
\end{equation}
The dependence on the ALH-coordinates $z_{j}$ and $\bar{z}_{j}$ is 
'hidden' in the functions $w_{k} = w_{k}( z,\bar{z} )$ 
(which are related to the old ones,
$u^{(1,3)}_{k}$, by $w_{k} \propto \sqrt{ v u^{(1)}_{k} / u^{(3)}_{k} }$). 
This dependence is given by equations (\ref{dark-disperion}) which lead to
\begin{equation}
  w_{k} = w_{k}^{(0)} \, 
  \exp\left\{ \sum_{a=1}^{\infty} \left(
    \mu_{k a} z_{a} + \mu_{k a}^{*} \bar{z}_{a} 
  \right) \right\}
\end{equation}
where $w_{k}^{(0)}$ are arbitrary constants and 
$\mu_{k a}^{*}$ are complex conjugate to $\mu_{k a}$ while the last ones  
are determined by 
\begin{equation}
  \exp\left\{ 2i \sum_{a=1}^{\infty} 
    \mu_{\kappa a} \frac{ \zeta^{a} }{ a } 
  \right\}
  = 
  \frac{ \tan\left[\frac{\omega}{2}+i\gamma_{\kappa}-\theta(\zeta)\right] }
       { \tan\left[\frac{\omega}{2}+i\gamma_{\kappa}\right] }.
\end{equation}
Applying the $\zeta \frac{d}{d\zeta}$ to the logarithm 
of the last equation one can conclude that the quantities $\mu_{k a}$ are 
the coefficients of the following Taylor series:
\begin{equation}
  \sum_{a=1}^{\infty} \mu_{\kappa a} \zeta^{a} =  
  \frac{ i }{ 2 } \; 
  \frac{ \sin 2\omega }{ \sin 2\theta_{\kappa} } \;
  \frac{ \zeta }{ R(\zeta) } \;
  \frac{ 1 }{ 1 - \zeta \exp\left( 2i\vartheta_{\kappa} \right) } 
\end{equation}
Now we have all necessary to write the final formula for the $N$-dark 
soliton solution of the ALH:
\begin{equation}
  q_{n}[N] = q_{vac} 
  \exp\left( i \sum\nolimits_{\kappa} \vartheta_{\kappa} \right) \; 
  \frac{ \Delta^{(+)}_{n} }{ \Delta^{(0)}_{n} }
\hspace{20mm}
  r_{n}[N] = q^{*}_{n}[N]
\end{equation}
with
\begin{eqnarray}
  \Delta^{(0)}_{n} & = & 
  \det\biggl| \; 
  \left\{ 
    w_{\kappa}      h_{\kappa}^{n+j-1} +
    w_{\kappa}^{-1} h_{\kappa}^{-n-j+1} 
  \right\}
  \exp\left( - i \vartheta_{\kappa}j \right)
   \; \biggr|_{ j,k = 1, ... , N }
\\[3mm]
  \Delta^{(+)}_{n} & = & 
  \det\biggl| \; 
  \left\{ 
    e^{ i\phi_{\kappa}} w_{\kappa}      h_{\kappa}^{n+j-1} +
    e^{-i\phi_{\kappa}} w_{\kappa}^{-1} h_{\kappa}^{-n-j+1} 
  \right\}
  \exp\left( - i \vartheta_{\kappa}j \right)
   \; \biggr|_{ j,k = 1, ... , N }
\end{eqnarray} 
Here $q_{vac}=\sqrt{\alpha} \, u$ is given by 
\begin{equation}
  q_{vac}(z, \bar{z}) = 
  \sqrt{\alpha} 
  \exp\left\{ 
    i \sum_{a=1}^{\infty}
    \lambda_{a} \left( z_{a} + \bar{z}_{a} \right) 
  \right\}
\end{equation}
where $\lambda_{a}$ are the coefficients of the Taylor series
\begin{equation}
  \sum_{a=1}^{\infty} \lambda_{a} \zeta^{a} =
  \frac{ R(\zeta) - \zeta - 1 }{ 2R(\zeta) }
\end{equation}
(which follows from (\ref{vac-evol})).

\section{Conclusion.}

The main idea of this work was to analyse the equations describing the 
BTs for the ALH, using the so-called functional representation, 
when an infinite number of the evolutionary equations are replaced using 
the Miwa's shifts with a few difference equations.
These equations contain much more information than any of the equations 
of the hierarchy taken alone and this gave us the possibility of 
\textit{solving} the BT equations. The main result of this work 
is \textit{explicit} (though formal) solution of the BT equations, 
i.e. \textit{explicit} form of the BTs. 
The fact that this can be done is not surprising and is already known from 
the general Sato's theory. For example, in \cite{AvM} one can find rather 
comprehensive discussion 
of how this method works in the context of the KP equation.
The solutions of the B\"{a}cklund equations 
obtained in this paper were constructed of 
four elementary transformations: $n \to n \pm 1$, $m \to m \pm 1$, 
Miwa's shifts and 
multiplication by the factor of the form $a^{m}b^{n}$. 
Each of these transformations is almost trivial. For example, if we apply 
them to the dark-soliton vacuum (\ref{dark-vacuum}), they coincide and do not 
change the quantities $q_{n}$ and $r_{n}$. In the case of the bright solitons 
the situation is even worse. Say, the transformation $m \to m \pm 1$ 
(which is one of the 'elementary BTs' of \cite{S}) to the vacuum tau-function 
(\ref{tau-vac-bright}) he will obtain that new tau-functions are given by 
$\widetilde\sigma_{n}=1$ and $\widetilde\tau_{n}=\widetilde\rho_{n}=0$ 
which means that both $\widetilde{q}_{n}$ and $\widetilde{r}_{n}$ are 
undefined: $\widetilde{q}_{n} = 1/0$, $\widetilde{r}_{n} = 0/0$.
The key ingredient of the BTs is the linear superposition of different 
$\mathbb{B}^{(k)}$ and $\overline{\mathbb{B}}^{(k)}$, the superposition which 
is natural for the approach based on the IST technique \cite{MS}.
Taking as an example the results of section \ref{sec-sol}, 
one could note that 
the 'soliton' exponents (\ref{bright-Q}) and (\ref{bright-R}) appear when we 
solve equations determining the coefficients $u^{(k)}$ and $\bar{u}^{(k)}$ 
of the linear combinations (\ref{bright-bta}) and (\ref{bright-btd}).

Another important question is the following one:
are the BTs of this paper the standard Darboux-B\"{a}cklund
transformations or something different? As it was said in the introduction, 
in the modern theory of the BTs, based on the IST approach, the BTs appear as 
the results of the Darboux transformations applied to the auxiliary linear 
problems. In this paper, we have used another approach 
(trying to minimize the use of the auxiliary 
-- intermediate, in some sense -- problem)
and after having 
rederived the B\"{a}cklund equations did not return to (\ref{zcr-sp}). 
Nevertheless, I would like to state here, 
without presenting explicit formulae describing the action of our BTs at the 
linear level, that BT discussed in the present work are indeed the 
well-known Darboux-B\"{a}cklund transformations of the soliton theory. 
As an illustration to this fact, one can consider the results of section 
\ref{sec-sol} where the BTs were acting as soliton-adding transformations, 
which is the most typical example of the Darboux-B\"{a}cklund scheme.

Finally, it should be noted that approach of the presented work 
(to use functional representation instead of the auxiliary linear problem)
is not so sensitive to the boundary conditions than one based on the 
IST and can be more easily modified to the case of other classes of solutions 
of the ALH (not only soliton ones but, e.g., quasiperiodic, Wronskian, and 
other) without necessity to elaborate from scratch 
the corresponding inverse scattering scheme.

\section*{Acknowledgements.}

I am grateful to the referees of this paper for careful reading the 
manuscript, correcting misprints, adding important references and 
making suggestions on improving the presentation of the results.

This work has been supported by grants: BFM2003-02832 
(Ministerio de Educaci\'on y Ciencia, Spain), 
PAI-05-001 (Consejer\'{\i}a de Educaci\'on y Ciencia de la Junta de 
Comunidades de Castilla-La Mancha, Spain).

\section*{Appendix: superposition of Miwa's shifts \label{app-miwa}}
\renewcommand{\theequation}{A.\arabic{equation}}


\newcommand{\tauNN}[2]{\tau^{m#1}_{n#2}}

\newcommand{\tauAN}[2]{
  \left( 
    \mathbb{E}_{\xi}\, 
    \tau^{m#1}_{n#2} 
  \right)
}

\newcommand{\tauNA}[2]{
  \left( 
    \mathbb{E}_{\eta}\, 
    \tau^{m#1}_{n#2} 
  \right)
}

\newcommand{\tauAA}[2]{
  \left( 
    \mathbb{E}_{\xi} 
    \mathbb{E}_{\eta}\, 
    \tau^{m#1}_{n#2} 
  \right)
}

\newcommand{\tauBN}[2]{
  \left( 
    \overline{\mathbb{E}}_{\xi}\, 
    \tau^{m#1}_{n#2} 
  \right)
}

\newcommand{\tauNB}[2]{
  \left( 
    \overline{\mathbb{E}}_{\eta}\, 
    \tau^{m#1}_{n#2} 
  \right)
}

\newcommand{\tauAB}[2]{
  \left( 
    \mathbb{E}_{\xi} \overline{\mathbb{E}}_{\eta}\, 
    \tau^{m#1}_{n#2} 
  \right)
}

\newcommand{\tauBB}[2]{
  \left( 
    \overline{\mathbb{E}}_{\xi} 
    \overline{\mathbb{E}}_{\eta}\, 
    \tau^{m#1}_{n#2} 
  \right)
}


Here, I would like to present formulae describing superposition of Miwa's 
shifts. Some of them were derived in \cite{V2002}, other can be obtained 
from those of \cite{V2002} by simple algebra. 

The superposition of the positive Miwa's shifts $\etp{}$ can be described by
\begin{eqnarray}
  a(\xi,\eta) \;\;  
  \tauNN{}{}  \tauAA{}{+1}   & = &
  \xi   \tauAN{}{+1}  \tauNA{}{}  
   -  
  \eta  \tauAN{}{}  \tauNA{}{+1} 
\label{app-1}
\\[4mm]
  a(\xi,\eta) \;  
  \tauNN{}{}  \tauAA{+1}{}   & = &
  \xi   \tauAN{+1}{}  \tauNA{}{}  
   -  
  \eta  \tauAN{}{}  \tauNA{+1}{} 
\\[4mm]
  a(\xi,\eta) \; 
  \tauNN{}{}  \tauAA{+1}{+1}   & = &
  \xi   \tauAN{+1}{+1}  \tauNA{}{}  
  -  
  \eta  \tauAN{}{}  \tauNA{+1}{+1} 
\label{app-3}
\end{eqnarray}
where $a(\xi,\eta)$ is a constant (with respect to the indices $m$ and $n$) 
which should be determined form the boundary conditions.
Note that only two of these equations are independent: the third one can be 
obtained by applying (\ref{alh-restr}). 
In the same manner one can produce a large number of similar formulae. 
Another type of superposition formulae for the positive Miwa's shifts can 
be presented as follows:
\begin{eqnarray}
  c(\xi,\eta) \; 
  \tauAN{}{}  \tauNA{}{}   & = &
  \tauNN{}{-1}  \tauAA{}{+1} 
  +  
  \tauNN{-1}{} \tauAA{+1}{} 
\\[4mm]
  c(\xi,\eta) \;  
  \tauAN{}{}  \tauNA{}{+1} & = &
  \tauNN{}{}  \tauAA{}{+1} 
  +  
  \xi \, \tauNN{-1}{} \tauAA{+1}{+1} 
\\[4mm]
  c(\xi,\eta) \; 
  \tauAN{+1}{} \tauNA{}{} & = &
  \tauNN{}{}   \tauAA{+1}{} 
  -
  \eta \, \tauNN{}{-1} \tauAA{+1}{+1} 
\end{eqnarray}
where $c(\xi,\eta)=(\xi-\eta)/a(\xi,\eta)$.

The superposition of the negative Miwa's shifts $\etn{}$ can be described by
\begin{eqnarray}
  \bar{a}(\xi,\eta) \;  
  \tauNN{}{+1} \tauBB{}{} & = &
  \xi  \tauBN{}{}   \tauNB{}{+1} 
  -  
  \eta \tauBN{}{+1} \tauNB{}{} 
\\[4mm]
  \bar{a}(\xi,\eta) \; 
  \tauNN{}{} \tauBB{+1}{} & = &
  \xi  \tauBN{+1}{} \tauNB{}{} 
  -  
  \eta \tauBN{}{}   \tauNB{+1}{} 
\\[4mm]
  \bar{a}(\xi,\eta) \;  
  \tauNN{}{+1} \tauBB{+1}{} & = &
  \xi \tauBN{+1}{} \tauNB{}{+1} 
  -  
  \eta \tauBN{}{+1}     \tauNB{+1}{} 
\end{eqnarray}
or by 
\begin{eqnarray}
  \bar{c}(\xi,\eta) \;  
  \tauBN{}{} \tauNB{}{} & = &
   \tauNN{}{+1} \tauBB{}{-1}  
  +  
  \tauNN{-1}{} \tauBB{+1}{} 
\\[4mm]
  \bar{c}(\xi,\eta) \; 
  \tauBN{}{} \tauNB{}{+1} & = &
  \tauNN{}{+1}   \tauBB{}{} 
  +  
  \eta \, \tauNN{-1}{+1} \tauBB{+1}{} 
\\[4mm]
  \bar{c}(\xi,\eta) \; 
  \tauBN{}{} \tauNB{+1}{} & = &
  \tauNN{}{} \tauBB{+1}{} 
  -  
  \xi \, \tauNN{}{+1} \tauBB{+1}{-1} 
\end{eqnarray}
where, again, $\bar{c}(\xi,\eta)=(\xi-\eta)/\bar{a}(\xi,\eta)$.

Finally, for the mixed products of $\etp{}$ and $\etn{}$ shifts one can get 
formulae such as 
\begin{eqnarray}
  e(\xi,\eta) \; 
  \tauNN{}{} \tauAB{}{} & = &
  \tauAN{}{} \tauNB{}{} 
  - 
  \xi\eta \tauAN{}{+1} \tauNB{}{-1} 
\\[4mm]
  e(\xi,\eta) \;  
  \tauNN{-1}{} \tauAB{+1}{} & = &
  \tauAN{}{} \tauNB{}{} 
  -  
  \tauAN{}{+1} \tauNB{}{-1} 
\end{eqnarray}
and
\begin{eqnarray}
  g(\xi,\eta) \; 
  \tauAN{}{} \tauNB{}{} & = &
  \tauNN{}{} \tauAB{}{} 
  -  
  \xi\eta \, \tauNN{-1}{} \tauAB{+1}{} 
\\[3mm]
  g(\xi,\eta) \;  
  \tauAN{}{+1} \tauNB{}{-1} & = &
  \tauNN{}{} \tauAB{}{} 
  -  
  \tauNN{-1}{} \tauAB{+1}{} 
\\[3mm]
  g(\xi,\eta) \; 
  \tauAN{-1}{+1} \tauNB{}{} & = &
  \tauNN{-1}{+1} \tauAB{}{} 
  -  
  \xi \, \tauNN{-1}{} \tauAB{}{+1} 
\\[3mm]
  g(\xi,\eta) \;  
  \tauAN{+1}{+1} \tauNB{}{} & = &
  \tauNN{}{} \tauAB{+1}{+1} 
  -  
  \eta \, \tauNN{}{+1} \tauAB{+1}{} 
\end{eqnarray}
with $e(\xi,\eta) g(\xi,\eta) = 1 - \xi\eta$.



\begin{thebibliography}{99}
\bibitem{AL1}
  M.J. Ablowitz and J.F. Ladik, 1975,
  Nonlinear differential-difference equations.
  {\it J.Math. Phys.}, {\bf 16}, 598--603.
\bibitem{AL2}
  M.J. Ablowitz  and J.F. Ladik, 1976,
  Nonlinear differential-difference equations and Fourier analysis.
  {\it J. Math. Phys.}, {\bf 17}, 1011--1018.
\bibitem{RS}
  C. Rogers and W.K. Schief, 2002,
  {\it B\"{a}cklund and Darboux transformations : 
       geometry and modern applications in soliton theory.}
  (Cambridge University Press, Cambridge).
\bibitem{MS}
  V.B. Matveev and M.A. Salle, 1991, 
  {\it Darboux Transformations and Solitons.} 
  (Springer-Verlag, Berlin Heidelberg New York). 
\bibitem{CM}
  A.Roy Chowdhurry and G. Mahato, 1983,
  A Darboux-B\"{a}cklund transformation associated with a discrete nonlinear
  Schr\"{o}dinger equation.
  {\it Lett.Math.Phys.}, {\bf 7}, 313--317.
\bibitem{GX}
  Geng Xianguo, 1989,
  Darboux transformation of the discrete Ablowitz-Ladik eigenvalue problem.
  {\it Acta Math.Sci.}, {\bf 9}, 21--26.
\bibitem{PBL}
  F. Pempineli, M. Boiti, J. Leon, 1995,
  B\"{a}cklund and Darboux transformations for the 
  Ablowitz-Ladik spectral problem.
  {\it Proc. of the first workshop "Nonlinear Physics, Theory and Experiment",
       Le Sirenuse, Gallipoli (Lecce), Italy, June 29 - July 7}, 261--268.
\bibitem{R}
  D.E. Rourke, 2004,
  Elementary B\"{a}cklund transformations for a discrete 
  Ablowitz-Ladik eigenvalue problem.
  {\it J. Phys. A}, {\bf 37}, 2693--2708.
\bibitem{AvM}
  M. Adler and P. van Moerbeke, 1994,
  Birkhoff strata, Baecklund transformations, and regularization of
  isospectral operators.
  {\it Adv.Math.}, {\bf 108}, 140--204.
\bibitem{AS}
  M.J. Ablowitz and H. Segur, 1981,
  {\it Solitons and the Inverse Scattering Transform}
  (SIAM, Philadelphia).
\bibitem{SDK}
  A. Seeger, H. Donth and A. Kochend\"{o}rfer, 1953, 
  Theorie der Versetzungen in eindimensionalen Atomreihen III. 
  Versetzungen, Eigenbewegungen und ihre Wechselwirkung.
  {\it Z. Phys.}, {\bf 134}, 173–-193.
\bibitem{V1998}
  V.E. Vekslerchik, 1998,
  Functional representation of the Ablowitz-Ladik hierarchy.
  {\it J.Phys.A}, {\bf 31}, 1087--1099.
\bibitem{V2002}
  V.E. Vekslerchik, 2002,
  Functional representation of the Ablowitz-Ladik hierarchy. II.
  {\it J. Nonlin. Math. Phys.}, {\bf 9}, 157-180.
\bibitem{S}
  T. Sadakane, 2003, 
  Ablowitz-Ladik hierarchy and two-component Toda lattice hierarchy.
  {\it J.Phys.A}, {\bf 36}, 87--97.
\end{thebibliography}
\end{document}